\documentclass[12pt,preprint]{aastex}
\usepackage{natbib}
\usepackage{graphicx}
\usepackage{epstopdf}
\usepackage{color}
\usepackage{float}
\usepackage{natbib}

\tightenlines

\newcommand     \PSF     {$\Psi\,$}        
\newcommand     \mPSF  {\Psi}              
\newcommand     \um     {$\mu$m$\,$}        
\newcommand     \mic     {$\mu$m$\,$}          
\newcommand     \Angstrom       {\,{\rm \AA}}
\newcommand     \asec {$^{\prime \prime}\,$}
\newcommand     \masec {^{\prime \prime}\,}

\newcommand     \beq    {\begin{equation}}
\newcommand     \beqa   {\begin{eqnarray}}

\newcommand     \eeq   {\end{equation}}
\newcommand     \eeqa   {\end{eqnarray}}

\newcommand     \gtsim  {\gtrsim}                

\newlength{\figwidth}
\newlength{\figwidthw}
\newlength{\figwidthww}
\newlength{\figwidthd}
\begin{document}

\title         {Common-Resolution Convolution Kernels for Space- and Ground-Based Telescopes}
\shorttitle{Common-Resolution Convolution Kernels}

\author{G. Aniano\altaffilmark{1},
        B.~T.~Draine\altaffilmark{1},
        K.D.Gordon\altaffilmark{2},
        K. Sandstrom\altaffilmark{3}}
\altaffiltext{1}{Princeton University Observatory, 
                 Peyton Hall, 
                 Princeton, NJ 08544,USA. }
             
\altaffiltext{2}{SpaceTelescope Science Institute, 
                      3700 San Martin Drive, 
                      Baltimore, MD 21218, USA. }
\altaffiltext{3}{Max-Planck Institut fur Astronomie, 
                  D-69117 Heidelberg, Germany.}
                  
 \email{ganiano@astro.princeton.edu}

\begin{abstract}

Multiwavelength study of extended astronomical objects requires combining images from instruments with differing point-spread functions (PSFs).
We describe the construction of convolution kernels that allow one to generate (multiwavelength) images with a common PSF, thus preserving the colors of the astronomical sources.
We generate convolution kernels for the cameras of the Spitzer Space Telescope, Herschel Space Observatory, Galaxy Evolution Explorer (GALEX), Wide-field Infrared Survey Explorer (WISE), ground-based optical telescopes (Moffat functions and sum of Gaussians), and Gaussian PSFs. 
Kernels for other telescopes including IRAS, AKARI, and Planck, are currently being constructed.
These kernels allow the study of the spectral energy distribution (SED) of extended objects, preserving the characteristic SED in each pixel.
The convolution kernels and the IDL packages used to construct and use them are made publicly available.

\end{abstract}


\clearpage
\section{\label{sec:intro}Introduction}

Spectral energy distribution studies of astronomical objects provide insight into the ongoing physical processes. 
In order to achieve a wide range of wavelengths, it is often necessary to combine observations from cameras with very different PSFs, in some cases with full width half maximum (FWHM) differing by factors of 100. 
Direct comparison (e.g., ratios) of images with structures on multiple angular/spatial scales obtained with different PSFs can result in unphysical intensity ratios (i.e., colors).
To preserve colors, intensity ratios should be calculated from images with a common PSF.
We therefore require convolution kernels that will transform the images taken with several instruments into a common PSF, so we can generate image cubes (i.e., a collection of images expressed in the same sky coordinates grid) in which each pixel corresponds to the same sky region for all the cameras used.

By ``camera'' we refer to the combination of telescope optics and physical detector, including the effect of atmospheric ``seeing'' if applicable. The PSF $\mPSF_j(x,y,x',y')$ of a camera $j$ gives the measured intensity at $(x,y)$ produced by a point source with unit flux at the point $(x',y')$, where we use the Cartesian coordinates $(x,y)$ to denote positions in a small region of the sky.
With this definition, the PSF has normalization
\beq
\label{eq_PSFnorm}
\int\int \mPSF_j(x,y,x',y')dxdy=1,
\eeq
for any source position $(x',y')$.

It is often possible to approximate the PSF (denoted as \PSF from now on) as constant across the useful field of view of the camera, so $\mPSF_j(x,y,x',y') = \mPSF_j(x-x',y-y')$.
The observed image $I_j(x,y)$ will then be the convolution of the source $S(x,y)$ with the PSF \PSF:
\beq
I_j(x,y)=\int\int S(x',y')\,\mPSF_j(x-x',y-y')dx'dy'
= \left(S \star \mPSF_j\right)(x,y).
\eeq

Clearly, given two cameras $A$ and $B$, with (different)  PSFs $\mPSF_{A}$ and $\mPSF_{B}$, the images obtained of an astronomical object will be different, even if the spectral response of the cameras were identical.
A convolution kernel is a tool that transforms the image observed by one camera into an image corresponding to the PSF of another camera.  

The convolution kernel $K\{A \Rightarrow B\}$ from camera $A$ to camera $B$ should satisfy
\beq
I_B(x,y)=\int\int I_A(x',y')\,K\{A \Rightarrow B\}(x-x',y-y')dx'dy'
\equiv \left(I_A\star K\{A \Rightarrow B\}\right)(x,y),
\label{ker:def}
\eeq
where $I_A$ and $I_B$ are the observed images by the cameras $A$ and $B$ respectively. 

The actual PSF of an instrument will show variations with the source color, variations along the field of view, changes over time, and deviations from rotational symmetry.
Full 2D characterization of a PSF is extremely challenging, and its extended wings are often not well determined. 
To take into account deviations of PSFs from rotational symmetry would require separate kernels $K\{A \Rightarrow B\}(\gamma)$ for each relative orientation $\gamma$ of cameras $A$ and $B$.
In the present study, the actual PSFs are close enough of having rotational symmetry that such additional complication is not justified (see \S 3 for a detailed study).
The current work assumes that $\mPSF_{A}$ and $\mPSF_{B}$ can be approximated by rotationally symmetric functions.

Using a different technique, \citet{Alard+Lupton_1998} presented a method
for finding optimal kernels to convolve (to a common resolution)
images of a sky region taken with a single camera under different
seeing conditions.
Using techniques similar to those used here,
\citet{Gordon+Engelbracht+Rieke+etal_2008} created kernels for the Infrared Array Camera (IRAC) and the Multiband Imaging Photometer for Spitzer (MIPS) cameras of Spitzer, and \citet{Sandstrom+Bolatto+Stanimirovic+etal_2009} created kernels for the Infrared Spectrograph (IRS) of Spitzer.
In the present work, we use the latest characterization of the PSFs of the cameras on the Spitzer Space Telescope (IRAC and MIPS), Herschel Space Observatory (Photocamera Array Camera and Spectrometer for Herschel (PACS) and Spectral and Photometric Imaging Receiver for Herschel (SPIRE)), Galaxy Evolution Explorer (GALEX) (FUV, NUV), and the Wide-field Infrared Survey Explorer (WISE) (W1-W4).
Additional PSFs (including those characterizing the IRS spectrograph on board Spitzer; the PACS spectrographs on board the Herschel Space Observatory; and all-sky images produced by IRAS, AKARI, and Planck) are currently being constructed and will be added to the kernel library.
We also include a family of analytical PSF profiles that are commonly used to model the PSFs for ground based telescopes.
We construct the set of kernels to transform from different instrumental PSFs into Gaussian PSFs, another form that is widely used.
We find an optimal Gaussian PSF for each camera: i.e., a Gaussian PSF that is sharp enough to capture the angular resolution of the camera, and wide enough to be robust against image artifacts.

Additional image processing (i.e., co-adding images\footnote{For example, for 2MASS co-added images the images undergo an additional smearing with a kernel whose size corresponds to a detector pixel.}) or different data reduction schemes (i.e., the Scanamorphos pipeline \citep{Roussel_2011} for the Herschel images) will alter the  PSF, and thus new kernels should be constructed using the effective PSF.

This article is organized as follows. In \S 2 we describe the generation of convolution kernels, in \S 3 we describe the PSF used, and in \S 4 we describe the kernel construction strategy. 
The performance of the kernels is examined in \S 5, and in \S 6 we describe a set of Gaussian kernels that are compatible with the different instruments.
In \S 7 we describe the kernel usage and show their performance on NGC 1097, and we summarize in \S 8.

All the kernels, IDL routines to use the kernels and IDL routines to make new kernels, along with detailed analysis of the generated kernels, are publicly available.\footnote{See http://www.astro.princeton.edu/$\sim$draine/Kernels.html. Kernels for additional cameras and updates will be included when new PSF characterizations become available.}

\section{Convolution Kernels}

Given two cameras $A$ and $B$, with PSFs $\mPSF_{A}$ and $\mPSF_{B}$, we seek $K\{A \Rightarrow B\}$ that fulfills equation
(\ref{ker:def}) for any astronomical source. Thus,
\beq
\left(S\star \mPSF_B\right)=I_B= \left(I_A\star K\{A \Rightarrow B\}\right)= \left(S\star \mPSF_A\star K\{A \Rightarrow B\}\right)
\eeq
 for any astronomical source $S$, so the convolution kernel must satisfy
\beq
\label{ker:real}
\mPSF_B=\left(\mPSF_A\star K\{A \Rightarrow B\}\right).
\eeq
With the normalization condition given by equation (\ref{eq_PSFnorm}), the kernel must have
\beq
 \label{eq_Kernorm}
\int\int K\{A \Rightarrow B\}(x,y)dxdy=1.
\eeq

We can easily invert equation (\ref{ker:real}) in Fourier space; taking the two-dimensional Fourier transform ($FT$) of equation (\ref{ker:real})\footnote{Computing the two-dimensional Fourier transform of rotationally symmetric functions is mathematically equivalent to making a decomposition in the one-parameter family of Bessel functions. However, the existence of the fast fourier transform (FFT) algorithm makes it numerically more efficient to perform the decomposition in the family of Fourier modes.} we obtain
\beq
FT\left(\mPSF_B \right) =FT\left(\mPSF_A\star K\{A \Rightarrow B\}\right) = FT\left(\mPSF_A\right)\times FT\left(K\{A \Rightarrow B\}\right).
\eeq
This can be inverted to obtain
\beq
\label{ker:Fourier}
K\{A \Rightarrow B\} = FT^{-1} \left( FT(\mPSF_{B})\times {1\over FT(\mPSF_{A})} \right),
\eeq
where $FT$ and $FT^{-1}$ stand for the Fourier transform and its inverse transformation respectively.

Equation (\ref{ker:Fourier}) provides a condition for the existence of such a kernel and a practical way of computing it.
We can see that a condition for the existence of a kernel is that the Fourier components for which $FT(\mPSF_{A}) = 0$ should satisfy  $FT(\mPSF_{B}) = 0$.
Informally speaking, this means that the PSF of camera $A$ must be narrower than the PSF of camera $B$.

For each camera $A$, we identify a high-frequency cutoff $k_{H,A}$ as the highest spatial frequency for which FT($\mPSF_A$) is appreciable by setting
\beq
FT(\mPSF_A)(k_{H,A})=5\times10^{-3} \times \mathrm{max}[FT(\mPSF_A)].
\eeq
The  cutoff frequency $k_{H,A}$ can be normalized as $ k_{H,A}=\kappa_A\times 2 \pi / \mathrm{FWHM}_A $. 
The values of $\kappa_A$ are in the range of 1.08 - 1.46 and are given in  Table \ref{tab_PSF_info}.

Equation (\ref{ker:Fourier}) also provides insight of a possible problem in computing kernels. 
PSF Fourier transforms do not have significant power at (spatial) frequencies above the  $ 2 \pi / \mathrm{FWHM}$.
The high-frequency components of the FT will be small, introducing large uncertainties when inverted.

A possible way of avoiding problems in the high frequency components of the kernel is to introduce a filter $f_A$ in the kernel construction:
\beq
\label{ker:Fourier_filter}
K\{A \Rightarrow B\} = FT^{-1} \left( FT(\mPSF_{B})\times {1\over FT(\mPSF_{A})} \times f_A \right),
\eeq
where $f_A $ is a suitable low-pass filter. 
Because this differs from equation (\ref{ker:Fourier}), it is clear that a kernel satisfying equation (\ref{ker:Fourier_filter}) will, in general, not be an exact solution to equation (\ref{ker:Fourier}).
However, we can expect that if the filter $f_A$ does not remove significant  power from the high frequency components of either $FT(\mPSF_A)$ or $FT(\mPSF_B)$, then the kernel computed will be a good approximate solution to equation (\ref{ker:Fourier}).

We use a filter $f_A$ of the form
\beq
\label{filter}
f_A (k)=
\left \{ \begin{array}{lcc} 
1                                                                                                                                       &\mathrm{ for }&k \le k_{L,A}\\
{1\over2}\times \left[1 + \cos \left(\pi \times {k - k_{L,A} \over k_{H,A}-k_{L,A}}\right) \right]  &\mathrm{ for }&k_{L,A}\le k \le k_{H,A}, \\
0                                                                                                                                       &\mathrm{ for }&k_{H,A}\le k
\end{array} \right.
\eeq
where we set $ k_{L,A} = 0.7 \times k_{H,A}$. 
Note that the cutoff frequency of our filter $f_A$ depends only on camera $A$, since small values of $FT(\mPSF_{B})$ have no negative impact in the kernel construction.
In principle, any smoothly varying function that is close to 1 in the low-frequency range and goes to zero in frequencies larger than $k_{H,A}$ should work as well. 
We have experimented using several smoothing functions and find that the particular form of $f_A (k)$ given by  equation (\ref{filter}), with these particular choices of $ k_{L,A}$ and $k_{H,A}$, gives excellent results. 
More details of the smoothing function effects can be found in \S 5.

If the cutoff frequencies associated with cameras $A$ and $B$ are such that $k_{H,B} \le  k_{H,A} \times 0.7$, then the filter $f_A$ has little effect and the resulting kernel from equation  (\ref{ker:Fourier_filter}) will satisfy equation (\ref{ker:def}) to a very good approximation.
In the regime $ k_{H,A} \times 0.7 \le k_{H,B} \le k_{H,A}$ the kernel $K\{A \Rightarrow B\}$ will depend on the exact form of the filter $f_A$ used. 
Most of the kernels in this regime have good performance; the performance of such ``filtered'' kernels will be evaluated in \S 5.
A limiting case is a kernel $K\{A \Rightarrow A\}$ that transforms a PSF into itself; it is the Fourier transform of the filter $f_A$.

In the cases of $k_{H,A} < k_{H,B}$ (convolving into narrower PSFs), use of the filter $f_A$ allows one to compute convolution kernels, but their performance can be poor. 
This will be further discussed subsequently.

\section{Instrumental point-spread functions}

We generate appropriate kernels for the measured PSFs of the cameras on Spitzer, Herschel, GALEX, and WISE and for certain analytical PSFs that are used in the literature:

IRAC.-- The Infrared Array Camera \citep{Fazio+Hora+Allen+etal_2004} has four infrared bands, centered at 3.6\um,
5.0\um, 5.8\um, and 8.0\um. We use the in-flight  extended point response functions (PRF).\footnote{In the current work we always use the full response of the optical systems including the camera effects, and for simplicity we will not make any further distinction between PSFs and PRFs} The core portion of the PSF was made with 300 observations of a calibration source, with different exposure times, combined into a high dynamic range image. Observations of the stars Vega, $\epsilon$Eridani, Fomalhaut, $\epsilon$Indi and Sirius were used in the construction of the extended wings of the PSF\footnote{The PSFs available at: http://ssc.spitzer.caltech.edu/}.

MIPS.-- The Multiband Imaging Photometer for Spitzer \citep{Rieke+Young+Engelbracht+etal_2004} has three photometric infrared bands, centered at 24\um,
70\um, and 160\um. 
Following \citet{Engelbracht+Blaylock+Su+etal_2007, Gordon+Engelbracht+Fadda+etal_2007, Stansberry+Gordon+Bhattacharya+etal_2007}, we generate theoretical PSFs for the MIPS cameras, assuming a blackbody source at T =  25K.
The PSFs are generated using the software STinyTim\footnote{STinyTim is available at: http://irsa.ipac.caltech.edu/data/SPITZER/docs/dataanalysistools/tools/contributed/general/stinytim/}
 in a 0.5\asec grid, and they are smoothed with a square kernel with sides of 4.5\asec, 13.5\asec, and 25.5\asec for the bands at 70\um, 100\um, and 160\um respectively.
The smoothing sizes correspond to 1.6, 1.35, and 1.8 times the camera detector pixel size, and they should cause the core of the theoretical PSF to be in close agreement with the calibration point-source images.

PACS.-- The Photocamera Array Camera and Spectrometer for Herschel \citep{Poglitsch+Waelkens+Geis+etal_2010} has three photometric infrared bands, centered at 70\um, 100\um and 160\um. 
We use the in-flight PSF  \citep{Geis+Lutz_2010, Lutz_2010, Muller+PaCS_ICC_2010}; the core was defined by observations of  the star $\alpha$ Tau and the asteroid Vesta, with extended wings reconstructed from (saturated) observations of Mars, Neptune, IK Tau, and the Red Rectangle.  The azimuthally averaged encircled 
energy fraction of the PSFs out to 1000\asec were obtained from the HCSS/HIPE software.

SPIRE.-- The Spectral and Photometric Imaging Receiver for Herschel \citep{Griffin+Abergel+Abreu+etal_2010} has three photometric far-infrared bands, centered at 250\um, 350\um and 500\um.\footnote{We use the in-flight 1.0\asec PSF maps from ftp://ftp.sciops.esa.int/pub/hsc-calibration/SPIRE/}

GALEX.--  The Galaxy Evolution Explorer \citep{Martin+Fanson+Schiminovich+etal_2005} has two ultraviolet bands, FUV (1350 - 1750 \Angstrom) and NUV (1750 - 2800 \Angstrom).\footnote{ 
In-flight PSFs are available at http://www.galex.caltech.edu/researcher/techdoc-ch5.html}

WISE.-- The Wide-field Infrared Survey Explorer \citep{Wright+Eisenhardt+Mainzer+etal_2010} has four photometric infrared bands, centered at 3.4\um, 4.6\um, 12\um, and 22\um.  
The PSF shape varies significantly over the focal plane due to distortion from the telescope optics; so a library of PSFs corresponding to a $9\times9$ grid of locations on the focal plane was determined. 
For each camera, we average the 81 different PSFs to generate a single PSF per camera.

As a way of measuring the departure of a PSF $\mPSF_j$ from rotational symmetry, we define an asymmetry parameter:
\beq
g_j \equiv \int\int \mid  \mPSF_j -C[\mPSF_j] \mid dxdy\\,
\label{def:g}
\eeq
where $C[\mPSF_j]$ is the azimuthally- averaged version of $\mPSF_j$ (i.e., $C[\mPSF_j]$  is a PSF with rotational symmetry and the same radial profile as $\mPSF_j$). 
The PSFs used have $g\lesssim 0.2$. 

\begin{figure*} 
\centering 
\includegraphics[width=17cm,height=21cm,clip=true,trim=0.0cm 0.0cm 0.0cm 0.0cm]
 {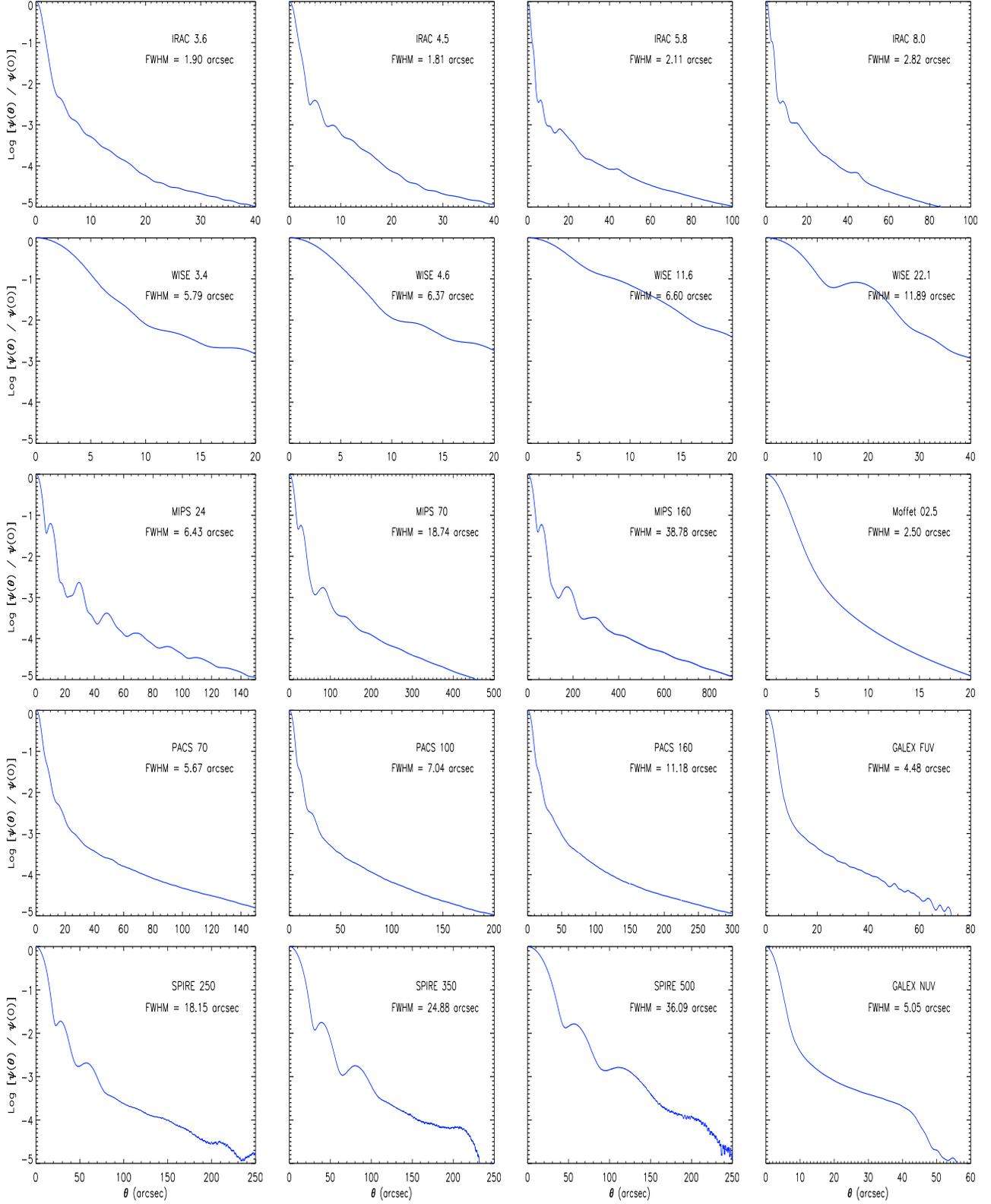} 
\caption{PSF radial profiles for the 20 cameras considered here (see text). All PSFs are shown out to $\mPSF(\theta)\approx 10^{-5} \mPSF(0)$, with the exception of WISE (W1-W4), for which we have data only down to $\approx 0.002 \mPSF(0)$}
\label{im_PSF_log}
\end{figure*}

\begin{figure*} 
\centering 
\includegraphics[width=17cm,height=21cm,clip=true,trim=0.0cm 0.0cm 0.0cm 0.0cm]
 {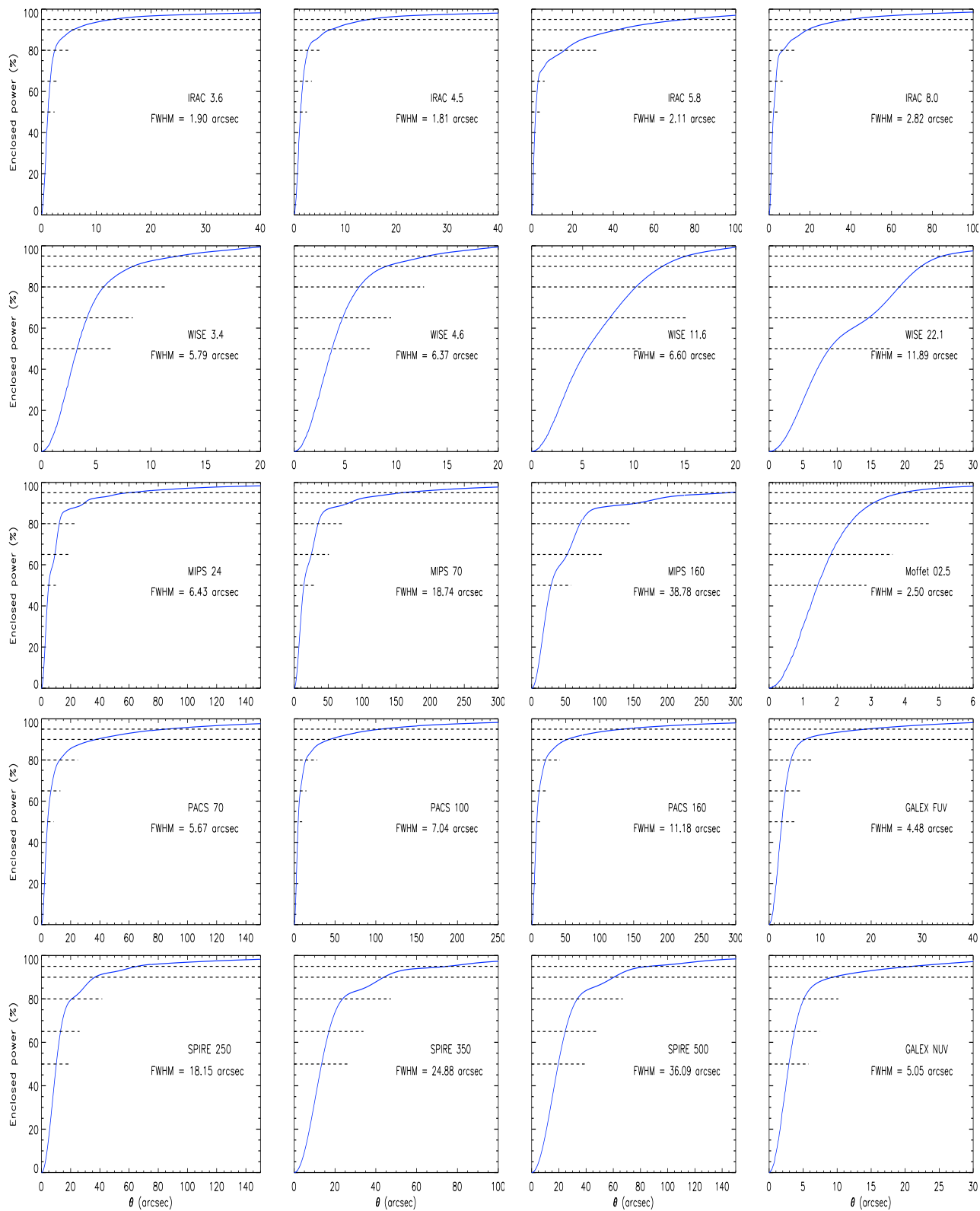} 
\caption{Fraction of the power enclosed by a circle of radius $\theta$  for the 20 cameras considered (see text).}
\label{im_PSF_int}
\end{figure*}

In Table \ref{tab_PSF_info} we have a summary of the properties of the different PSFs: the camera Rayleigh diffraction angle, the PSF FWHM, the radius containing 99\% of the PSF energy, the (normalized) high-frequency cutoff $\kappa_A$ used in the filter $f_A$, and the anisotropy parameter $g$. 
The PSF radial profiles [out to $\mPSF(\theta)\approx 10^{-5} \mPSF(0)$], and enclosed power are plotted in Figures \ref{im_PSF_log} and \ref{im_PSF_int}. 

For each PSF the radii containing 25\%, 50\%, 65\%, 80\%, 90\%, 95\%,  98\%, 99\%, 99.5\%, and 99.9\% of the total power are given in Table \ref{tab_PSF_energy_radii}. 
Table \ref{tab_PSF_energy_perc} gives the enclosed power for selected radii.

\begin{table*}[h] 
\caption{Basic Instrument Information} 
\label{tab_PSF_info}
\footnotesize
\centering 
\begin{tabular}{|l|ccccc|} 
\hline \hline 
                 & Rayleigh diff. & Measured  & 99\% of energy & $\kappa$$^b$& Asymmetry \\
Camera           &   limit$^a$ (\asec)   &   FWHM  (\asec)  &    radius  (\asec)    & &  $g$$^c$  \\
\hline \hline 
IRAC 3.6\mic     &           1.04 &       1.90 &      62.52 &       1.29 &       0.16 \\ 
IRAC 4.5\mic     &           1.31 &       1.81 &      64.46 &       1.26 &       0.17 \\ 
IRAC 5.8\mic     &           1.68 &       2.11 &     133.55 &       1.20 &       0.19 \\ 
IRAC 8.0\mic     &           2.30 &       2.82 &     114.20 &       1.19 &       0.18 \\ 
\hline
MIPS 24\mic      &           6.93 &       6.43 &     224.53 &       1.05 &       0.08 \\ 
MIPS 70\mic      &          20.90 &      18.74 &     461.44 &       1.12 &       0.05 \\ 
MIPS 160\mic     &          45.62 &      38.78 &     678.77 &       1.10 &       0.05 \\ 
\hline
PACS 70\mic      &           5.11 &       5.67 &     249.81 &       1.23 &       0.20 \\ 
PACS 100\mic     &           7.28 &       7.04 &     350.63 &       1.19 &       0.20 \\ 
PACS 160\mic     &          11.70 &      11.18 &     417.36 &       1.21 &       0.20 \\ 
\hline
SPIRE 250\mic    &          17.93 &      18.15 &     205.07 &       1.16 &       0.19 \\ 
SPIRE 350\mic    &          25.16 &      24.88 &     192.47 &       1.15 &       0.18 \\ 
SPIRE 500\mic    &          36.22 &      36.09 &     198.43 &       1.16 &       0.19 \\ 
\hline
GALEX FUV        &           0.08 &       4.48 &      50.28 &       1.26 &       0.07 \\ 
GALEX NUV        &           0.11 &       5.05 &      39.56 &       1.32 &       0.05 \\ 
\hline
WISE 3.35\mic    &           2.11 &       5.79 &      19.10 &       1.20 &       0.17 \\ 
WISE 4.60\mic    &           2.89 &       6.37 &      19.08 &       1.33 &       0.13 \\ 
WISE 11.56\mic   &           7.27 &       6.60 &      19.56 &       1.23 &       0.12 \\ 
WISE 22.1\mic    &          13.90 &      11.89 &      35.15 &       1.05 &       0.07 \\ 
\hline
Gauss 12\asec    &                &     12.00 &       15.41    &      1.33  &     0.0   \\ 
Gauss 20\asec    &                &     20.00 &       25.68    &      1.33  &     0.0   \\ 
Gauss 23\asec    &                &     23.00 &       29.53    &      1.33  &     0.0   \\ 
Gauss 28\asec    &                &     28.00 &       35.95    &      1.33  &     0.0   \\ 
Gauss 40\asec    &                &     40.00 &       51.33    &      1.33  &     0.0   \\ 
Gauss 50\asec    &                &     50.00 &       64.05    &      1.33  &     0.0   \\ 
\hline 
BiGauss 0.5\asec &                &      0.50 &        0.90    &      1.37  &     0.0   \\ 
BiGauss 1.0\asec &                &      1.00 &        1.79    &      1.37  &     0.0   \\ 
BiGauss 1.5\asec &                &      1.50 &        2.69    &      1.37  &     0.0   \\ 
BiGauss 2.0\asec &                &      2.00 &        3.57    &      1.37  &     0.0   \\ 
BiGauss 2.5\asec &                &      2.50 &        4.44    &      1.37  &     0.0   \\ 
\hline 
Moffat 0.5\asec  &                &      0.50 &        1.39    &      1.46  &     0.0   \\ 
Moffat 1.0\asec  &                &      1.00 &        2.77    &      1.46  &     0.0   \\ 
Moffat 1.5\asec  &                &      1.50 &        4.12    &      1.46  &     0.0   \\ 
Moffat 2.0\asec  &                &      2.00 &        5.45    &      1.46  &     0.0   \\ 
Moffat 2.5\asec  &                &      2.50 &        6.74    &      1.46  &     0.0   \\ 
\hline \hline 
\multicolumn{6}{l}{$^a$ We take the Rayleigh diffraction limit as 1.22 $\times$ [central $\lambda$] / [telescope diameter].}\\
\multicolumn{6}{l}{$^b$ We define $\kappa$ as $(k_{H}\times \mathrm{FWHM}_A)/(2\pi) $ where $k_{H}$ is the high-frequency cutoff (see text for details).}\\
\multicolumn{6}{l}{$^c$ The parameter $g$ is a measure of the departure of a PSF from rotational symmetry [see eq.(\ref{def:g})].}\\
\end{tabular} 
\end{table*} 
   
\begin{table*}[h] 
\caption{PSF Enclosed Energy Percent (\%) in Selected Circular Apertures} 
\label{tab_PSF_energy_radii}
\footnotesize
\centering 
\begin{tabular}{|@{}c@{}|ccccccccccccccc|} 
\hline \hline 
      &  \multicolumn{15}{|c|}{Radii} \\
     &  \multicolumn{15}{|c|}{(\asec)}   \\
Camera &  2.5 & 5.0  & 7.5 & 10  & 12.5 & 15  & 17.5 & 20  & 25  & 30  & 40  & 50 & 60 & 90 & 120   \\
\hline \hline 
      IRAC 3.6\mic&  80.7 &     88.7 &     91.8 &     93.6 &     94.8 &     95.7 &     96.3 &     96.7 &     97.2 &     97.6 &     98.2 &     98.6 &     98.9 &     99.5 &     99.8 \\
      IRAC 4.5\mic&  78.1 &     86.7 &     90.3 &     92.5 &     94.0 &     95.1 &     95.9 &     96.5 &     97.1 &     97.5 &     98.1 &     98.6 &     98.9 &     99.5 &     99.8 \\
      IRAC 5.8\mic&  57.6 &     70.5 &     74.2 &     76.0 &     77.6 &     79.2 &     81.2 &     82.9 &     85.4 &     87.0 &     89.5 &     91.7 &     93.2 &     96.3 &     98.3 \\
      IRAC 8.0\mic&  53.9 &     78.0 &     80.8 &     84.0 &     85.9 &     87.6 &     89.3 &     90.5 &     92.3 &     93.5 &     95.1 &     96.2 &     96.9 &     98.3 &     99.2 \\
\hline 
       MIPS 24\mic&  21.6 &     50.5 &     59.4 &     69.1 &     81.2 &     85.5 &     86.5 &     87.2 &     88.3 &     90.4 &     92.7 &     94.0 &     95.0 &     96.7 &     97.8 \\
       MIPS 70\mic&  3.00 &     11.2 &     22.5 &     34.2 &     44.3 &     51.5 &     56.1 &     59.1 &     64.4 &     71.8 &     83.9 &     87.1 &     88.0 &     91.3 &     93.4 \\
      MIPS 160\mic&  0.73 &     2.88 &     6.32 &     10.8 &     16.1 &     21.8 &     27.8 &     33.5 &     43.8 &     51.3 &     58.9 &     63.4 &     70.4 &     86.9 &     88.7 \\
\hline 
       PACS 70\mic&  26.8 &     56.8 &     69.3 &     76.6 &     80.0 &     82.2 &     84.1 &     85.6 &     87.3 &     88.6 &     90.4 &     91.8 &     92.9 &     95.3 &     96.7 \\
      PACS 100\mic&  20.4 &     51.3 &     65.0 &     71.8 &     77.6 &     80.8 &     82.5 &     83.7 &     86.2 &     87.7 &     89.5 &     90.8 &     91.9 &     94.2 &     95.6 \\
      PACS 160\mic&  8.89 &     29.3 &     49.0 &     61.2 &     67.8 &     72.3 &     76.2 &     79.3 &     82.8 &     84.7 &     87.6 &     89.5 &     90.8 &     93.1 &     94.5 \\
\hline 
     SPIRE 250\mic&  4.39 &     16.3 &     32.4 &     48.9 &     62.6 &     71.8 &     77.0 &     79.5 &     82.2 &     85.9 &     91.0 &     92.5 &     94.0 &     96.5 &     97.5 \\
     SPIRE 350\mic&  2.48 &     9.56 &     20.1 &     32.7 &     45.6 &     57.2 &     66.9 &     73.9 &     81.2 &     83.5 &     87.8 &     92.5 &     94.0 &     96.6 &     97.9 \\
     SPIRE 500\mic&  1.17 &     4.62 &     10.1 &     17.2 &     25.5 &     34.3 &     43.2 &     51.7 &     66.0 &     75.8 &     83.9 &     86.5 &     89.9 &     95.2 &     97.1 \\
\hline 
     GALEX FUV    &  50.2 &     84.9 &     90.5 &     92.2 &     93.2 &     94.0 &     94.6 &     95.2 &     96.2 &     97.0 &     98.2 &     99.0 &     99.5 &      100 &      100 \\
     GALEX NUV    &  40.7 &     79.3 &     88.1 &     90.5 &     91.9 &     93.0 &     93.9 &     94.7 &     96.1 &     97.2 &     99.1 &     99.8 &     99.9 &      100 &      100 \\
\hline 
     WISE 3.35\mic&  34.5 &     74.8 &     87.7 &     92.6 &     95.0 &     96.9 &     98.2 &     99.4 &      100 &      100 &      100 &      100 &      100 &      100 &      100 \\
     WISE 4.60\mic&  28.4 &     68.0 &     85.9 &     91.5 &     94.4 &     96.7 &     98.2 &     99.4 &      100 &      100 &      100 &      100 &      100 &      100 &      100 \\
    WISE 11.56\mic&  18.6 &     46.1 &     64.0 &     78.9 &     89.1 &     94.8 &     97.5 &     99.3 &      100 &      100 &      100 &      100 &      100 &      100 &      100 \\
     WISE 22.1\mic&  7.36 &     24.9 &     42.8 &     54.3 &     60.4 &     66.0 &     74.0 &     83.1 &     94.5 &     97.6 &     99.7 &      100 &      100 &      100 &      100 \\
\hline 
Gauss 12\asec   &  11.3 & 38.2 & 66.2 & 85.5 & 95.1 & 98.7 & 99.8 &  100 &  100 &  100 & 100  &  100 & 100  &  100 &  100 \\
Gauss 20\asec   &  4.24 & 15.9 & 32.3 & 50.0 & 66.2 & 79.0 & 88.1 & 93.8 & 98.7 & 99.8 &  100 &  100 & 100  &  100 &  100 \\
Gauss 23\asec   &  3.22 & 12.3 & 25.5 & 40.8 & 55.9 & 69.3 & 79.9 & 87.7 & 96.3 & 99.1 &  100 &  100 & 100  &  100 &  100 \\
Gauss 28\asec   &  2.18 & 8.46 & 18.0 & 29.8 & 42.5 & 54.9 & 66.2 & 75.7 & 89.1 & 95.9 & 99.7 &  100 &  100 &  100 &  100 \\
Gauss 40\asec   &  1.08 & 4.24 & 9.29 & 15.9 & 23.7 & 32.3 & 41.2 & 50.0 & 66.2 & 79.0 & 93.8 & 98.7 & 99.8 &  100 &  100 \\
Gauss 50\asec   &  0.69 & 2.73 & 6.05 & 10.5 & 15.9 & 22.1 & 28.8 & 35.8 & 50.0 & 63.2 & 83.1 & 93.8 & 98.2 &  100 &  100 \\
\hline 
BiGauss 0.5\asec&   100 &      100 &      100 &      100  &  100 &  100 &  100 &  100 &  100 &  100 &  100 &  100 &  100 &  100 &  100 \\
BiGauss 1.0\asec&  99.9 &      100 &      100 &      100  &  100 &  100 &  100 &  100 &  100 &  100 &  100 &  100 &  100 &  100 &  100 \\
BiGauss 1.5\asec&  98.6 &      100 &      100 &      100  &  100 &  100 &  100 &  100 &  100 &  100 &  100 &  100 &  100 &  100 &  100 \\
BiGauss 2.0\asec&  95.7 &     99.9 &      100 &      100  &  100 &  100 &  100 &  100 &  100 &  100 &  100 &  100 &  100 &  100 &  100 \\
BiGauss 2.5\asec&  89.9 &     99.5 &      100 &      100  &  100 &  100 &  100 &  100 &  100 &  100 &  100 &  100 &  100 &  100 &  100 \\
\hline 
Moffat 0.5\asec &        100 &      100 &      100 &      100 &  100 &  100 &  100 &  100 &  100 &  100 &  100 &  100 &  100 &  100 &  100 \\
Moffat 1.0\asec &       98.7 &      100 &      100 &      100 &  100 &  100 &  100 &  100 &  100 &  100 &  100 &  100 &  100 &  100 &  100 \\
Moffat 1.5\asec &       96.1 &     99.5 &      100 &      100 &  100 &  100 &  100 &  100 &  100 &  100 &  100 &  100 &  100 &  100 &  100 \\
Moffat 2.0\asec &       90.9 &     98.7 &     99.7 &      100 &  100 &  100 &  100 &  100 &  100 &  100 &  100 &  100 &  100 &  100 &  100 \\
Moffat 2.5\asec &       83.0 &     97.7 &     99.3 &     99.8 &  100 &  100 &  100 &  100 &  100 &  100 &  100 &  100 &  100 &  100 &  100 \\
\hline \hline 
\end{tabular} 
\end{table*} 

\begin{table*}[h] 
\caption{Radii (\asec ) Enclosing Selected Percents of the Total Power} 
\label{tab_PSF_energy_perc}
\footnotesize
\centering 
\begin{tabular}{|c|cccccccccc|} 
\hline \hline 
         &  \multicolumn{10}{|c|}{Percent}\\
         &  \multicolumn{10}{|c|}{ (\%)}\\
Camera &  25 & 50 & 65& 80 & 90 & 95 & 98 & 99 & 99.5 & 99.9 \\
\hline \hline 
 IRAC 3.6\mic &     0.74 &     1.19 &     1.60 &     2.44 &     5.83 &     12.9 &     36.0 &     62.5 &     88.1 &      132 \\
 IRAC 4.5\mic &     0.75 &     1.24 &     1.73 &     2.68 &     7.06 &     14.7 &     37.9 &     64.5 &     90.8 &      134 \\
 IRAC 5.8\mic &     0.99 &     1.97 &     3.17 &     16.0 &     42.3 &     75.3 &      114 &      134 &      145 &      154 \\
 IRAC 8.0\mic &     1.16 &     2.21 &     3.32 &     6.70 &     18.9 &     39.5 &     82.9 &      114 &      134 &      152 \\
\hline 
  MIPS 24\mic &     2.73 &     4.93 &     9.18 &     12.2 &     29.4 &     60.4 &      130 &      225 &      368 &      738 \\
  MIPS 70\mic &     8.02 &     14.4 &     25.5 &     35.6 &     80.2 &      158 &      318 &      461 &      628 &      894 \\
 MIPS 160\mic &     16.3 &     29.0 &     52.8 &     71.9 &      155 &      287 &      518 &      679 &      802 &      950 \\
\hline 
  PACS 70\mic &     2.41 &     4.23 &     6.47 &     12.5 &     37.5 &     85.3 &      165 &      250 &      378 &      712 \\
 PACS 100\mic &     2.83 &     4.87 &     7.50 &     14.1 &     43.5 &      105 &      227 &      351 &      477 &      711 \\
 PACS 160\mic &     4.50 &     7.66 &     11.3 &     20.8 &     53.3 &      133 &      294 &      417 &      524 &      712 \\
\hline 
SPIRE 250\mic &     6.39 &     10.2 &     13.1 &     20.8 &     36.6 &     66.1 &      138 &      205 &      382 &      488 \\
SPIRE 350\mic &     8.49 &     13.4 &     17.0 &     23.7 &     43.9 &     75.0 &      122 &      192 &      397 &      499 \\
SPIRE 500\mic &     12.4 &     19.5 &     24.6 &     33.5 &     60.3 &     85.9 &      137 &      198 &      411 &      511 \\
\hline 
GALEX FUV     &     1.59 &     2.49 &     3.17 &     4.27 &     7.05 &     19.0 &     38.3 &     50.3 &     59.7 &     75.1 \\
GALEX NUV     &     1.82 &     2.90 &     3.71 &     5.08 &     9.34 &     20.9 &     33.9 &     39.6 &     43.4 &     58.7 \\
\hline 
WISE 3.35\mic &     2.03 &     3.26 &     4.17 &     5.67 &     8.36 &     12.5 &     17.2 &     19.1 &     20.2 &     21.4 \\
WISE 4.60\mic &     2.29 &     3.69 &     4.75 &     6.36 &     9.00 &     13.0 &     17.2 &     19.1 &     20.2 &     21.4 \\
WISE 11.56\mic&     3.02 &     5.48 &     7.66 &     10.2 &     12.8 &     15.1 &     18.1 &     19.6 &     20.5 &     21.5 \\
WISE 22.1\mic &     5.01 &     8.87 &     14.6 &     19.1 &     22.3 &     25.4 &     31.2 &     35.1 &     38.5 &     42.1 \\
\hline 
Gauss 12\asec   & 3.86 & 6.00 & 7.38 & 9.14 & 10.9 & 12.5 & 14.2 & 15.4 & 16.5 & 18.5 \\
Gauss 20\asec   & 6.44 &10.00 & 12.3 & 15.2 & 18.2 & 20.8 & 23.7 & 25.7 & 27.5 & 30.9 \\
Gauss 23\asec   & 7.41 & 11.5 & 14.1 & 17.5 & 20.9 & 23.9 & 27.3 & 29.5 & 31.6 & 35.5 \\
Gauss 28\asec   & 9.02 & 14.0 & 17.2 & 21.3 & 25.5 & 29.1 & 33.2 & 36.0 & 38.5 & 43.2 \\
Gauss 40\asec   & 12.9 & 20.0 & 24.6 & 30.5 & 36.4 & 41.5 & 47.4 & 51.3 & 54.9 & 61.5 \\
Gauss 50\asec   & 16.1 & 25.0 & 30.8 & 38.1 & 45.5 & 51.9 & 59.2 & 64.1 & 68.4 & 76.3 \\
\hline 
BiGauss 0.5\asec&  0.17 &     0.26 &     0.32 &     0.41 &     0.50 &     0.60 &     0.76 &     0.90 &     1.02 &     1.25 \\
BiGauss 1.0\asec&  0.33 &     0.52 &     0.64 &     0.81 &     1.00 &     1.20 &     1.53 &     1.79 &     2.04 &     2.50 \\
BiGauss 1.5\asec&  0.50 &     0.78 &     0.97 &     1.22 &     1.50 &     1.80 &     2.29 &     2.69 &     3.05 &     3.71 \\
BiGauss 2.0\asec&  0.66 &     1.04 &     1.29 &     1.62 &     2.00 &     2.40 &     3.04 &     3.57 &     4.04 &     4.86 \\
BiGauss 2.5\asec&  0.83 &     1.30 &     1.61 &     2.03 &     2.50 &     3.00 &     3.79 &     4.44 &     5.01 &     5.96 \\
\hline 
Moffat 0.5\asec &   0.18 &     0.29 &     0.36 &     0.47 &     0.60 &     0.76 &     1.07 &     1.39 &     1.72 &     2.26 \\
Moffat 1.0\asec &   0.36 &     0.57 &     0.72 &     0.94 &     1.21 &     1.53 &     2.14 &     2.77 &     3.43 &     4.50 \\
Moffat 1.5\asec &   0.54 &     0.86 &     1.08 &     1.41 &     1.81 &     2.29 &     3.19 &     4.12 &     5.09 &     6.70 \\
Moffat 2.0\asec &   0.71 &     1.14 &     1.44 &     1.88 &     2.42 &     3.05 &     4.23 &     5.45 &     6.69 &     8.90 \\
Moffat 2.5\asec &   0.89 &     1.43 &     1.81 &     2.34 &     3.02 &     3.80 &     5.27 &     6.74 &     8.29 &     11.1 \\
\hline \hline 
\end{tabular} 
\end{table*} 

\clearpage

\clearpage

For convenience, we add several families of PSFs that are often used for ground-based optical and radio telescopes. For each analytical profile, we generate the kernels for a range of FWHM values. We consider the following analytical profiles:

Gaussians.-- We use Gaussian PSF of the form
\beq
\mPSF (\theta) ={1\over 2 \,\pi\, \sigma^2}  \exp \left({-\theta ^2 \over 2\sigma^2}\right),
\eeq
where the FWHM  $=2\times\sigma \, \sqrt{2 \ln 2}$. 
We generate kernels with 5\asec $<$ FWHM$<$65\asec.

Optical (SDSS).-- In the SDSS survey, it is found that a good approximation to the telescope PSF is given by the sum of two Gaussians. 
The two components have relative weights of 0.9 and 0.1, and the FWHM of the second component is twice that of the first. 
We use a family of PSFs of the form:
\beq
\mPSF (\theta) = 0.9{1\over 2 \,\pi\, \sigma^2}  \exp \left({-\theta ^2 \over 2\sigma^2}\right) +
                 0.1 {1\over 2 \,\pi\,  (2\sigma)^2}  \exp \left({-\theta ^2 \over 2(2\sigma)^2}\right).
\eeq
where the FWHM  $=1.01354 \times2\times\sigma \, \sqrt{2 \ln 2}$. 
We generate kernels with FWHM = 0.5, 1.0, 1.5, 2.0, and 2.5\asec.

Optical (General).-- \citet{Moffat_1969} proposed PSFs of the form:
\beq
M_\beta(\theta) = {(\beta - 1)(2^{1/\beta}-1)\over \pi \,{\theta_0}^2} \left[ 1+ (2^{1/\beta}-1)\left({ r \over \theta_0}\right)^2\right]^{-\beta},
\eeq
where $\beta$ is a parameter, and $\mathrm{FWHM} = 2\,\theta_0$, 
Following \citet{Racine_1996} we use PSFs of the form:
\beq
\mPSF (r) = 0.8\times M_{7}(\theta)+ 0.2 \times M_{2}(\theta),
\eeq
where the same $\theta_0$ is used in $M_7$ and $M_2$. 
We generate kernels with FWHM = $2\,\theta_0$ = 0.5, 1.0, 1.5, 2.0, and 2.5\asec.

\section{Kernel Generation}

Generation of the convolution kernels was accomplished as follows:

\subsection{Input the PSF and correct for missing data.}

When an input PSF image has missing data pixels (like the current SPIRE PSFs), we iteratively estimate the missing values.

We start by replacing the missing data pixels by a value of 0 in the original image.
We compute a smoothed image by convolving it with a (normalized) Gaussian kernel $\propto \exp[-(\theta/2\theta_0)^2]$, with $\theta_0$ equal to 2 pixels. 
We replace the original image missing data pixels by the value they have in the convolved image (the original data are not altered).
We iterate the convolution and replacement steps 5 times.
The resulting image has all the missing data points replaced by a smooth interpolation. 
This technique produces robust results, even if we have missing data in a patch of a few contiguous pixels.

\subsection{Resample the PSFs.}

Each PSF comes in a grid of different pixel size. 
We transform each PSF into a grid of a common pixel size of 0.20\asec using the IDL procedure {\it congrid}, using the \textit{cubic convolution} interpolation method with a parameter of -0.5. 
The 0.20\asec pixel size capture all the details on the instrumental and Gaussian PSFs. 
We also pad with 0 the resulting images into an odd-sized square array if needed.

We use a grid of a common pixel size of 0.10\asec to regenerate the kernels from optical PSFs into IRAC cameras.

\subsection{Center the PSFs.}

To determine the image center, we smooth the image with a 5 pixel radius circular kernel, and locate the image maximum. 
In some PSFs the maximum value is achieved over a (small) ring. 
To avoid possible misidentification of the real image center, we take the centroid of all the pixels that satisfy:
\beq
{\mathrm{max}[\mPSF]-\mPSF(x,y) \over \mathrm{max}[\mPSF]} \le 5\times 10^{-4}.
\eeq

\subsection{Circularize the PSFs.}
 
In order to make a rotationally symmetric PSF, we average over $2^{14}$ rotations of the image through angles $\beta_n=n\times360^\circ / 2^{14}$ for $n=1,2,3,...,2^{14}=16,384$, producing a PSF image that is invariant under rotations of any angle that is a multiple of $360^\circ /2^{14} = 0.022^\circ $ (i.e., is as rotationally symmetric as we can numerically expect).

Computing $2^{14}$ rotations naively would be extremely time-consuming, but the final result can be in fact computed performing only 14 rotations, as follows.

We start by rotating \PSF by an angle $\theta_1=360^\circ / 2^{1}=180^\circ$, producing an image $R_1$, and computing their average 
$\overline{R_1}= {1\over2}\times[\mPSF + R_1]$.
Clearly, $\overline{R_1}$ is now invariant under rotations of $\theta_1=360^\circ / 2^{1}=180^\circ$.

We continue this procedure iteratively.  We rotate $R_1$ by an angle $\theta_2=360^\circ / 2^{2}=90^\circ$, producing an image $R_2$,
and set  $\overline{R_2}= {1\over2}\times[\overline{R_1} + R_2]$.
$\overline{R_2}$ is invariant under rotations of $\theta_1=360^\circ / 2^{1}=180^\circ$ and $\theta_2=360^\circ / 2^{2}=90^\circ$; i.e. it is invariant under rotations of any angle that is a multiple of $360^\circ/2^{2} = 90^\circ $. 

We iterate this procedure 14 times; the last average calculated image $\overline{R_{14}}$ is the desired rotationally symmetric PSF.

We further set to 0 all the pixels that lie outside the largest circle included in the square image, since those regions would correspond to areas with partial image coverage. If there are pixels with (very small) negative values (due to the noise in the original PSFs) we set them to 0.

In order to have a more stable algorithm, the previous rotations are performed in reverse order (i.e., starting with the smallest angles).

All the remaining steps in the kernel construction should preserve the rotational symmetry in the images.
A way of estimating the noise induced by some steps (e.g., computing Fourier transform) is to compute the departure from rotational symmetry in the resulting image. 
Circularizing helps to reduce numerical noise, and will be performed after every step that can potentially decrease the image quality. 
When a circularization is performed to a rotationally symmetric image, the asymmetry parameter $g$ of the resulting image is smaller than 0.0008 (i.e., the circularization procedure itself induces very small departures from rotational symmetry).

\subsection{Resize the PSF.}

We trim (or pad with 0) all the PSFs into a common grid, to be able to compute all the convolution kernels using only one Fourier transform per PSF. We choose a grid size that is large enough to contain most of the power in each PSF.
We also optimize its size to make the FFT algorithm as efficient as possible (minimal sum of prime factors).
The adopted grid size is $729 \masec \times 729 \masec$, giving an image size of $3645 \times 3645$ pixels.

\subsection{Compute the Fourier transform of the PSF  $FT(\mPSF)$.}

We use an efficient FFT algorithm.
Since the PSFs are invariant under reflections, $\vec{x} \leftrightarrow -\vec{x}$, their Fourier transform should be real. We impose this real condition to reduce the numerical noise introduced by the double-precision FFT algorithm.

\subsection{Circularize the $FT(\mPSF)$.}

Using the procedure as before, we circularize the $FT(\mPSF)$. 
In principle, they should already be rotationally symmetric, but numerical noise in the FFT algorithm makes them slightly non rotationally symmetric. 
  
\subsection{Filter the $FT(\mPSF)$.}
 
We filter the highest frequencies in each $FT(\mPSF)$. We use a filter $\phi$ of the form
\beq
\label{filter2}
\phi (k)=
\left \{ \begin{array}{lcc} 
1                                                                                                                                       &\mathrm{ for }&k \le k_{\alpha}\\
 \exp \left[ -\left(1.8249 \times {k - k_{\alpha} \over k_{\beta}-k_{\alpha}}\right)^4 \right]  &\mathrm{ for }&k_{\alpha}\le k \le k_{\beta} ,\\
0                                                                                                                                       &\mathrm{ for }&k_{\beta}\le k
\end{array} \right.
\eeq
where we set $ k_{\alpha} = 0.9 \times k_{\beta}$. The factor 1.8249 is chosen so that $\phi(0.5\times(k_{\beta}+k_{\alpha}))=0.5$. 
For each camera we choose $k_{\beta}=4\times (2\pi / \mathrm{FWHM})$.
We tested several filter functions and found that the particular form given by equation (\ref{filter2}) provided the best results.
Each PSF has structure at spatial wavelengths comparable with the FWHM, so the Fourier components with frequencies much higher than this cannot be important. 
The Fourier components removed by this filter were mainly introduced by the original image resizing algorithm. 
In the following discussion, we let $FT_\phi (\mPSF)=\phi \times FT(\mPSF)$.

\subsection{Invert the $FT(\mPSF)$.}

We evaluate $1/FT_\phi(\mPSF)$ at the points where $FT_\phi(\mPSF) \ne 0$.  
We set $1/FT_\phi(\mPSF)=0$ at the remaining points (which will be filtered soon).

\subsection{Compute the $FT$ of the filtered kernel.}

We compute the $FT$ of the filtered kernel using the filter $f_A$ from equation (\ref{filter}):

\beq
FT(K\{A \Rightarrow B\}) = FT_\phi(\mPSF_{B}) \times \frac{f_A}{FT_\phi(\mPSF_A)}
\label{eq:fil_FT}
\eeq

for all the appropriate combinations $(A,B)$.

\subsection{Compute the kernels.}

We compute the inverse Fourier transform to the previously calculated $FT(K)$. 
We again impose the condition that $K$ must be real.

\subsection{Circularize the kernels.}    

Using the procedure as before, we circularize the kernels. 
Again, they should already be rotationally symmetric. 
Numerical noise in the inverse FFT algorithm makes them slightly asymmetric, but this is easily corrected. 

\subsection{Resample  the kernels.}          

All the computed kernels are given in a grid of a common pixel size of 0.20\asec, but will be needed in grids of different pixel sizes. 
Again, we resample the kernels using the IDL procedure {\it congrid}, using the \textit{cubic convolution} interpolation method with a parameter of -0.5.

\subsection{Final trim of the kernels.}       

We trim each kernel to a smaller size (to speed up further convolution) such that it contains 99.9\% of the total kernel energy. 
Moreover, we use a square grid with an odd number of pixels so that the kernel peaks in a single central pixel.

\subsection{Circularize the final Kernels.}       

We finish the kernels by circularizing them again, to remove the small noise introduced in the resampling process.

All previous calculations were done in double precision to reduce numerical noise. 

\section{Kernel Performance}   

For each generated kernel, we compute $\mPSF_{A} \star K\{A \Rightarrow B\}$ the convolution of $\mPSF_{A}$ and $K\{A \Rightarrow B\}$, and compare it with $\mPSF_{B}$.\footnote{It can be easily shown that the radial profile of $(\mPSF_{A} \star K\{A \Rightarrow B\})$ is the same whether $\mPSF_{A}$ is rotationally symmetric or not, so for simplicity, we use the circularized version of $\mPSF_{A}$ in the comparison.}
For a perfect kernel, both quantities should coincide at all radii.

\begin{figure*}[h] 
\centering 
\begin{tabular}{lr} 
\includegraphics[width=8cm,height = 8.0cm,clip=true,trim=0.1cm 0.0cm 0.0cm 1.2cm]
{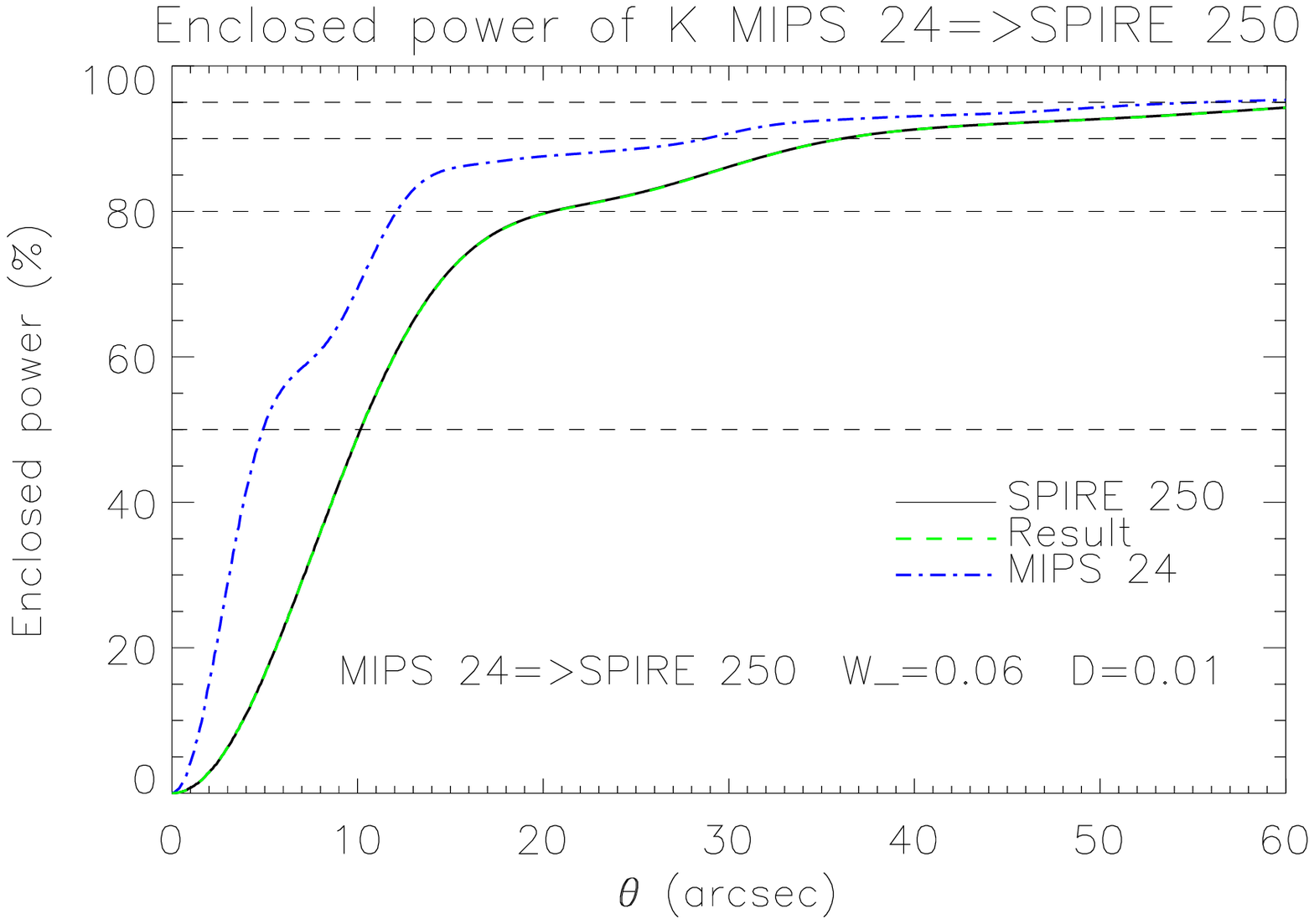} &
\includegraphics[width=8cm,height = 8.0cm,clip=true,trim=0.1cm 0.0cm 0.0cm 1.2cm]
{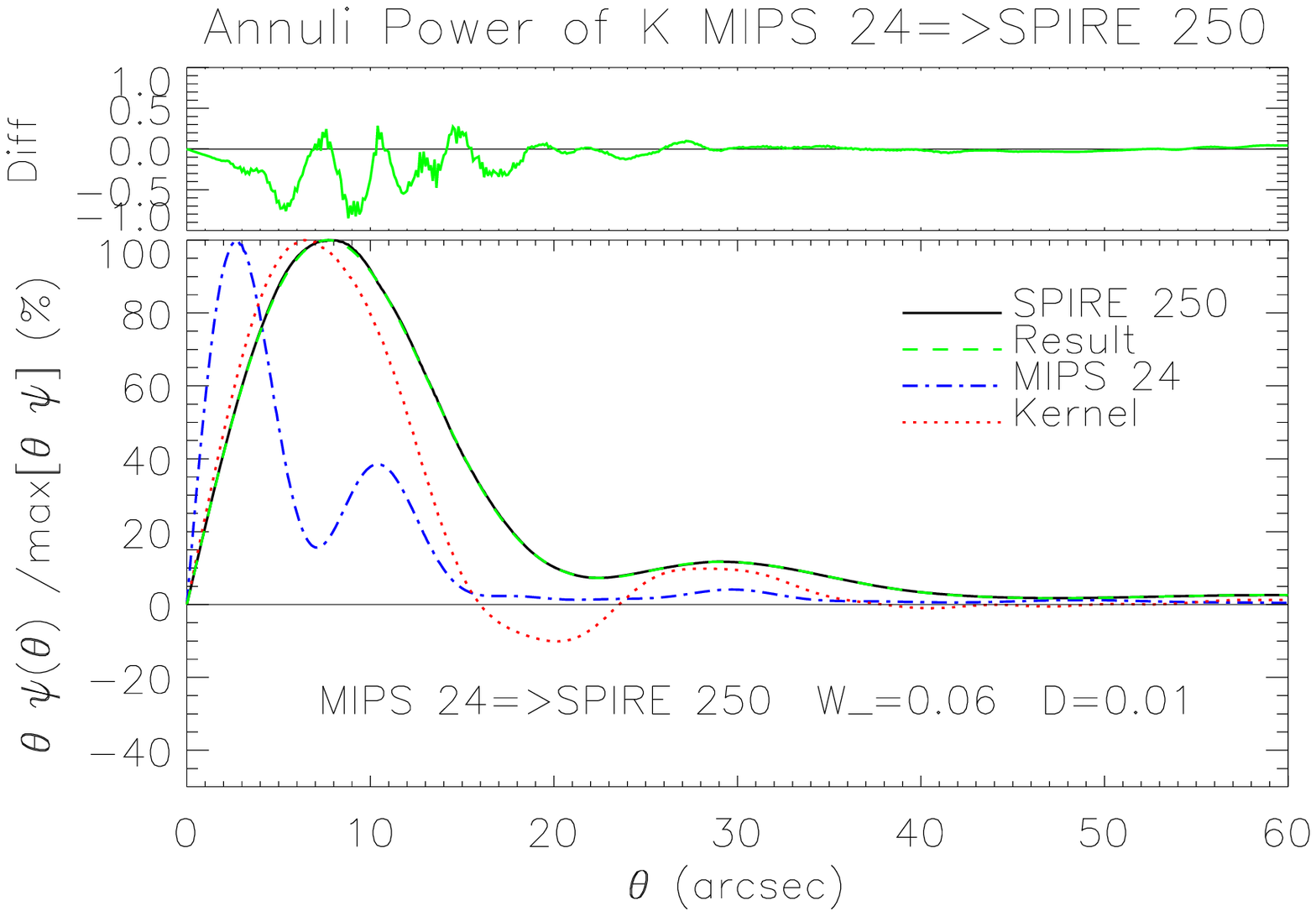} \\
\end{tabular} 
\caption{Performance of the filtered kernel $K\{\mathrm{M}24 \Rightarrow\mathrm{S}250\}$. $W_{-}$ is the integral of the negative values of the kernel, and $D$ is the integral of the absolute value of the difference between the target PSF and the PSF reproduced by the kernel [see eq. (\ref{eq:D})].
See the electronic edition of the PASP for a color version of this figure.}
\label{fig:perf}
\end{figure*}

Figure \ref{fig:perf} shows the analysis of $K\{\mathrm{M}24 \Rightarrow\mathrm{S}250\}$.\footnote{ 
In our PSF and kernel notation, we will abbreviate I, M, P, S, GAL, and W for IRAC, MIPS, PACS, SPIRE, GALEX, and WISE respectively, and will omit the \mic symbol.}
This kernel shows good behavior: it transforms from a camera with  $\mathrm{FWHM}_{\mathrm{M}24 }$  = 6.5\asec into a camera with $\mathrm{FWHM}_{\mathrm{S}250 }$ = 18.2\asec. 
This kernel essentially spreads the energy of the core of MIPS 24\mic PSF into a larger area.  
The filter $f_{\mathrm{M}24}$ has no effect on the construction of this kernel, because $FT(\mPSF_B)\approx0$ for $k> 0.7 \times k_{H,A}$

The left panel of Figure \ref{fig:perf} compares the integrated power of the PSFs. 
It includes $\mPSF_{\mathrm{M}24 }$ (dot-dashed line), $\mPSF_{\mathrm{S}250}$ (solid line), and $\mPSF_{\mathrm{M}24} \star K\{\mathrm{M}24 \Rightarrow\mathrm{S}250\}$ (dashed line). 
For an ideal kernel the solid line and the dashed line should coincide. 
The departures in this case are very small.

The power per unit radius is proportional to $\theta\Psi(\theta)$, or $\theta K(\theta)$. 
The right panel of Figure \ref{fig:perf} shows $\theta\Psi$ and $\theta K$ (normalized to the maximum value). 
The lower part of the right panel includes four traces: $\theta \times \mPSF_{\mathrm{M}24 }$ (dot-dash line), $\theta \times \mPSF_{\mathrm{S}250 }$ (solid line), $\theta \times (\mPSF_{\mathrm{M}24} \star K\{\mathrm{M}24 \Rightarrow\mathrm{S}250\})$ (dashed line) and $\theta \times K\{\mathrm{M}24 \Rightarrow\mathrm{S}250\}$ (dotted line) for visualization of the kernel behavior. For an ideal kernel the solid line and the dashed line should coincide. 
The upper part of the right panel is a plot of the difference between $\theta \times \mPSF_{\mathrm{S}250 }$ and $\theta \times (\mPSF_{\mathrm{M}24} \star K\{\mathrm{M}24 \Rightarrow\mathrm{S}250\})$. 
For an ideal kernel this graph should be 0. 
For this example ($K\{\mathrm{M}24 \Rightarrow\mathrm{S}250\}$), the $\mPSF_{\mathrm{M}24} \star K\{\mathrm{M}24 \Rightarrow\mathrm{S}250\}$ reproduces the  SPIRE 250\mic PSF to within $0.01\times\mPSF_{\mathrm{S}250}(0)$.

\begin{figure}[h] 
\centering 
\includegraphics[width=8.6cm,height = 8.0cm,clip=true,trim=0.0cm 0.0cm 0.0cm 1.2cm]
{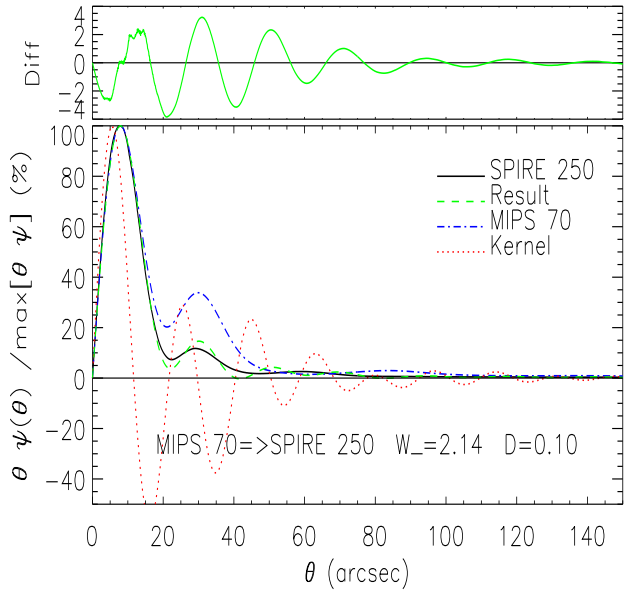} 
\caption{Performance of the kernel $K\{\mathrm{M}70 \Rightarrow\mathrm{S}250\}$.
Because of the large value $W_{-}=2.14$, convolution MIPS70\um $\Rightarrow$ SPIRE250\um is not recommended.
See the electronic edition of the PASP for a color version of this figure.}
\label{fig:perf2}
\end{figure}

\begin{figure}[h] 
\centering 
\includegraphics[width=8.6cm,height = 8.0cm,clip=true,trim=0.0cm 0.0cm 0.0cm 1.2cm]
{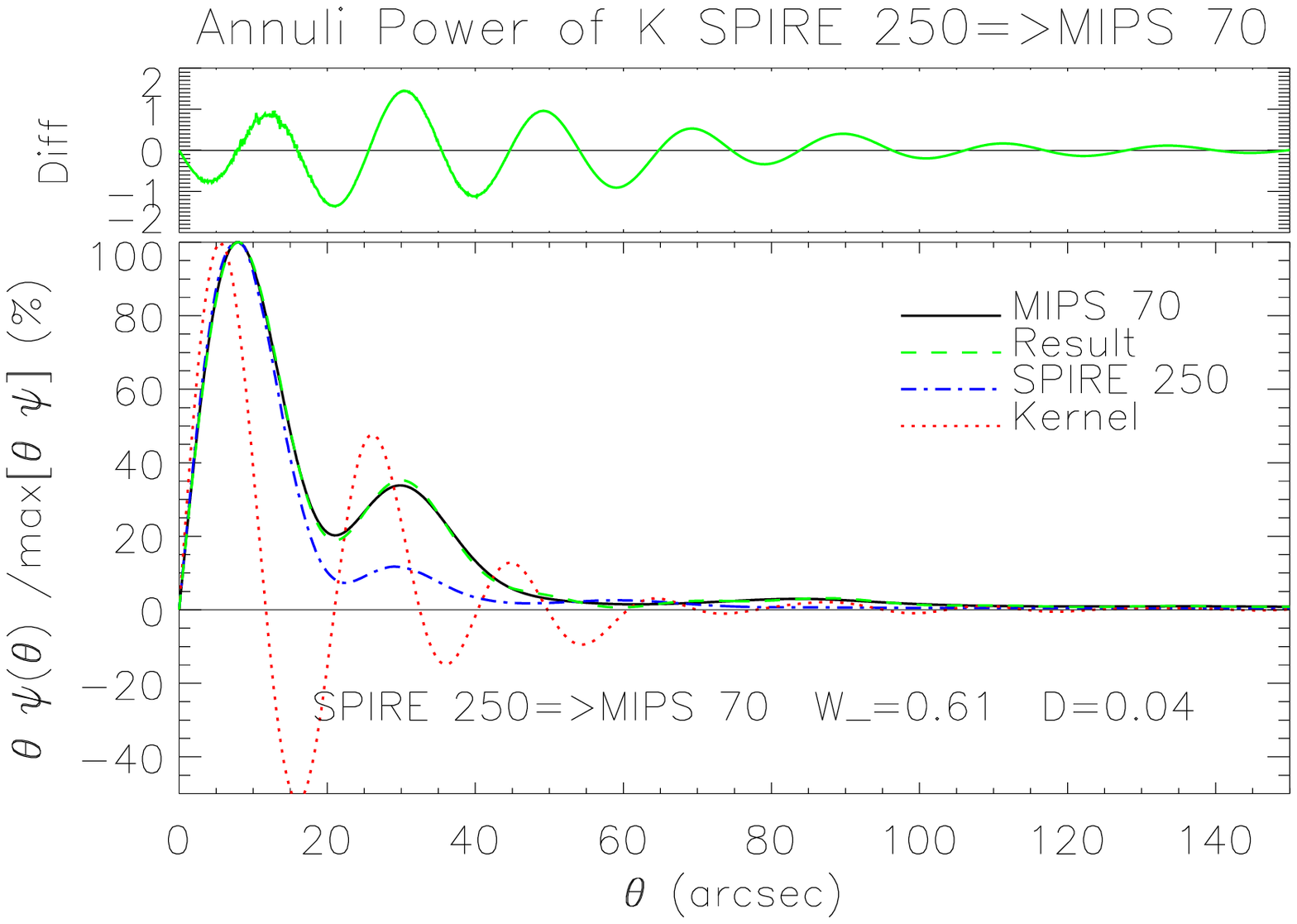} 
\caption{Performance of the kernel $K\{\mathrm{S}250 \Rightarrow\mathrm{M}70\}$.
With $W_{-}=0.61$, this kernel is safe to use. 
$\mPSF_{\mathrm{S}250} \star K\{\mathrm{S}250 \Rightarrow \mathrm{M}70\}$ deviates from $\mPSF_{\mathrm{M}70}$ less than 2\%.
See the electronic edition of the PASP for a color version of this figure.}
\label{fig:perf3}
\end{figure}

In Figures \ref{fig:perf2} and \ref{fig:perf3} we analyze the kernels $K\{\mathrm{M}70 \Rightarrow\mathrm{S}250\}$ and $K\{\mathrm{S}250 \Rightarrow \mathrm{M}70\}$. 
Their construction is more challenging since both cameras have similar FWHM: $\mathrm{FWHM}_{\mathrm{M}70} = 18.7\masec$ and $\mathrm{FWHM}_{\mathrm{S}250} = 18.2\masec$. 
These kernels have to redistribute the energy within the core and Airy rings of the PSFs. 
The plotted quantities are similar to those in the right panel of Figure \ref{fig:perf} for $K\{\mathrm{M}24 \Rightarrow\mathrm{S}250\}$. 
In this case, the kernels still perform well, but they have large areas with negative values.

One measure of kernel performance is its accuracy in redistribution of PSF power. We define
\beq
\label{eq:D}
D=\int\int \mid  \mPSF_{B} -K\{A \Rightarrow B\}\star \mPSF_{A}\mid dxdy\\.
\eeq
A kernel with perfect performance will have $D=0$, and normalization of the PSFs requires $D\le2$. 
$D$ measures how much flux has not been redistributed correctly.
Good kernels have small $D$ values: $D(K\{\mathrm{M}24 \Rightarrow\mathrm{S}250\}) = 0.011$. 
In Table \ref{tab_Ker_D} we give $D$ for the kernels constructed.

\begin{table*}[h] 
\caption{$D$ Value for Constructed Kernels} 
\label{tab_Ker_D} 
\footnotesize
\centering 
\begin{tabular}{|c|c@{$\,\,\,$}c@{$\,\,\,$}c | c@{$\,\,\,$}c@{$\,\,\,$}c | c@{$\,\,\,$}c@{$\,\,\,$}c | c@{$\,\,\,$}c | c@{$\,\,\,$}c@{$\,\,\,$}c@{$\,\,\,$}c|} 
\hline \hline 
 & \multicolumn{15}{|c|}{To} \\
 \cline{2-16}
       & M    & M   & M   & P   & P   & P    & S    & S    & S    & GAL  & GAL    & W     & W     & W     & W     \\ 
From             & 24    & 70   & 160  & 70   & 100  & 160   & 250   & 350   & 500   & FUV  & NUV    & 3.4   & 4.6   & 11.6  & 22.1  \\ 
\hline 
I3.6          & 0.065  & 0.029 & 0.037 & 0.051 & 0.047 & 0.031 & 0.025 & 0.022 & 0.018 & 0.006 & 0.005 & 0.005 & 0.005 & 0.007 & 0.008 \\ 
I4.5          & 0.065  & 0.029 & 0.037 & 0.051 & 0.047 & 0.031 & 0.025 & 0.022 & 0.018 & 0.006 & 0.005 & 0.005 & 0.005 & 0.007 & 0.008 \\
I5.8          & 0.064  & 0.031 & 0.048 & 0.047 & 0.045 & 0.031 & 0.029 & 0.027 & 0.018 & 0.012 & 0.010 & 0.012 & 0.012 & 0.013 & 0.012 \\
I8.0          & 0.064  & 0.030 & 0.047 & 0.047 & 0.046 & 0.032 & 0.026 & 0.024 & 0.018 & 0.012 & 0.007 & 0.007 & 0.007 & 0.009 & 0.009 \\
\hline 
M24           & 0.091  & 0.018 & 0.048 & 0.213 & 0.068 & 0.015 & 0.011 & 0.013 & 0.017 &   NC  & 0.352 & 0.225 & 0.122 & 0.098 & 0.027 \\
M70           &   NC   & 0.055 & 0.038 &   NC  &   NC  &   NC  & 0.102 & 0.015 & 0.014 &   NC  &   NC  &   NC  &   NC  &   NC  &   NC  \\
M160          &   NC   &   NC  & 0.064 &   NC  &   NC  &   NC  &   NC  &   NC  & 0.169 &   NC  &   NC  &   NC  &   NC  &   NC  &   NC  \\
\hline 
P70           & 0.006  & 0.019 & 0.049 & 0.045 & 0.010 & 0.014 & 0.011 & 0.014 & 0.017 & 0.227 & 0.113 & 0.064 & 0.045 & 0.027 & 0.027 \\
P100          & 0.072  & 0.014 & 0.045 & 0.188 & 0.050 & 0.010 & 0.011 & 0.014 & 0.015 &   NC  &   NC  & 0.202 & 0.111 & 0.087 & 0.028 \\
P160          &   NC   & 0.009 & 0.040 &   NC  &   NC  & 0.044 & 0.013 & 0.014 & 0.015 &   NC  &   NC  &   NC  &   NC  &   NC  & 0.032 \\
\hline 
S250          &   NC   & 0.037 & 0.045 &   NC  &   NC  &   NC  & 0.058 & 0.009 & 0.014 &   NC  &   NC  &   NC  &   NC  &   NC  &   NC  \\
S350          &   NC   & 0.300 & 0.043 &   NC  &   NC  &   NC  &   NC  & 0.060 & 0.012 &   NC  &   NC  &   NC  &   NC  &   NC  &   NC  \\
S500          &   NC   &   NC  & 0.042 &   NC  &   NC  &   NC  &   NC  &   NC  & 0.055 &   NC  &   NC  &   NC  &   NC  &   NC  &   NC  \\
\hline 
FUV             & 0.065  & 0.030 & 0.036 & 0.049 & 0.046 & 0.032 & 0.025 & 0.024 & 0.018 & 0.046 & 0.017 & 0.007 & 0.006 & 0.007 & 0.007 \\
NUV             & 0.065  & 0.030 & 0.037 & 0.051 & 0.046 & 0.032 & 0.027 & 0.024 & 0.018 & 0.071 & 0.028 & 0.008 & 0.006 & 0.006 & 0.007 \\
\hline 
W3.4           & 0.066  & 0.029 & 0.037 & 0.080 & 0.048 & 0.032 & 0.026 & 0.023 & 0.018 & 0.252 & 0.122 & 0.053 & 0.034 & 0.017 & 0.007 \\
W4.6           & 0.066  & 0.029 & 0.037 & 0.079 & 0.047 & 0.032 & 0.026 & 0.023 & 0.018 &   NC  & 0.118 & 0.051 & 0.033 & 0.016 & 0.006 \\
W11.6          & 0.074  & 0.028 & 0.037 & 0.117 & 0.056 & 0.031 & 0.026 & 0.023 & 0.018 &   NC  & 0.206 & 0.109 & 0.063 & 0.040 & 0.005 \\
W22.1          &   NC   & 0.028 & 0.036 &   NC  &   NC  & 0.163 & 0.028 & 0.023 & 0.018 &   NC  &   NC  &   NC  &   NC  &   NC  & 0.094 \\
\hline \hline 
\multicolumn{16}{l}{Notes.-- $D$ is the integral of the absolute value of the difference between} \\
\multicolumn{16}{l}{ a target PSF and the PSF reproduced by the kernel [see eq. (\ref{eq:D})].} \\
\multicolumn{16}{l}{ We are abbreviating I, M, P, S, GAL, and W }\\
\multicolumn{16}{l}{for IRAC, MIPS, PACS, SPIRE, GALEX, and WISE respectively.}\\
\multicolumn{16}{l}{  NC stands for not computed: the kernel performance would be too poor.}\\
\end{tabular} 
\end{table*} 

A second quantitative measure of kernel performance is obtained by studying its negative values. 
We define
\beq
\label{eq:W}
W_{\pm} = {1\over2} \int\int \left( \mid K\{A \Rightarrow B\} \mid \pm K\{A \Rightarrow B\} \right) dxdy.
\eeq

Flux conservation requires that $W_{+} = 1 + W_{-}$. 
In general, kernels will have $W_->0$. 
Well-behaved kernels have small $W_{-}$ values: $W_{-}(K\{\mathrm{M}24 \Rightarrow\mathrm{S}250\}) = 0.07$. 
The integral of $\mid K\{A \Rightarrow B\}\mid $ is $[1+2W_{-}]$, so a kernel with a large value of $W_{-}$ could potentially amplify image artifacts.
Additionally, a kernel with large $W_{-}$ can generate areas of negative flux near point sources if nonlinearities are present, or if the background levels were subtracted incorrectly.
Table \ref{tab_Ker_W} lists the $W_{-}$ values for the kernels constructed.
The values in Table \ref{tab_Ker_W}  were computed numerically. 
Due to finite grid resolution the numerical values may be off by a few percent in some cases. 
This can be seen from the $W_{-}$ values computed for the self-kernels. 
The self-kernels are simply the Fourier transform of the filter $f_{A}$, and $W_{-}$  should therefore be the same (1.15) in all cases.
However, in Table \ref{tab_Ker_W} the $W_{-}$ for the smallest PSFs are larger than 1.15 by as much as 0.03 (e.g., 1.18 for PACS 70\um).

\begin{table*}[h] 
\caption{$W_{-}$ Value for Constructed Kernels} 
\label{tab_Ker_W} 
\footnotesize
\centering 
\begin{tabular}{|c|c@{$\,\,\,$}c@{$\,\,\,$}c | c@{$\,\,\,$}c@{$\,\,\,$}c | c@{$\,\,\,$}c@{$\,\,\,$}c | c@{$\,\,\,$}c | c@{$\,\,\,$}c@{$\,\,\,$}c@{$\,\,\,$}c|} 
\hline \hline 
& \multicolumn{15}{|c|}{To} \\
 \cline{2-16}
       & M    & M   & M   & P   & P   & P    & S    & S    & S    & GAL  & GAL    & W     & W     & W     & W     \\ 
From             & 24    & 70   & 160  & 70   & 100  & 160   & 250   & 350   & 500   & FUV  & NUV    & 3.4   & 4.6   & 11.6  & 22.1  \\ 
\hline 
I3.6           &  0.00 & 0.00 & 0.00 & 0.00 & 0.00 & 0.00  & 0.00  & 0.00  & 0.00  & 0.06  & 0.03  & 0.05  & 0.05  & 0.05  & 0.02 \\ 
I4.5           &  0.01 & 0.00 & 0.00 & 0.00 & 0.00 & 0.00  & 0.00  & 0.00  & 0.00  & 0.07  & 0.03  & 0.05  & 0.05  & 0.05  & 0.02 \\
I5.8           &  0.15 & 0.03 & 0.02 & 0.07 & 0.07 & 0.03  & 0.05  & 0.05  & 0.02  & 0.29  & 0.23  & 0.25  & 0.24  & 0.23  & 0.15 \\
I8.0           &  0.11 & 0.00 & 0.00 & 0.01 & 0.01 & 0.00  & 0.00  & 0.01  & 0.00  & 0.23  & 0.11  & 0.14  & 0.14  & 0.12  & 0.06 \\
\hline 
M24            &  1.17 & 0.02 & 0.00 & 2.85 & 0.75 & 0.16  & 0.06  & 0.04  & 0.01  &  NC   & 5.14  & 3.28  & 1.83  & 1.11  & 0.32 \\
M70            &   NC  & 1.15 & 0.20 &  NC  &  NC  &  NC   & 2.14  & 0.66  & 0.34  &  NC   &  NC   &  NC   &  NC   &  NC   &  NC  \\
M160           &   NC  &  NC  & 1.14 &  NC  &  NC  &  NC   &  NC   &  NC   & 2.81  &  NC   &  NC   &  NC   &  NC   &  NC   &  NC  \\
\hline 
P70            &  0.83 & 0.01 & 0.01 & 1.18 & 0.32 & 0.00  & 0.05  & 0.05  & 0.02  & 5.74  & 2.68  & 1.48  & 0.77  & 0.48  & 0.16 \\
P100           &  2.16 & 0.01 & 0.01 & 4.29 & 1.17 & 0.04  & 0.07  & 0.06  & 0.03  &  NC   &  NC   & 4.82  & 2.56  & 1.75  & 0.24 \\
P160           &   NC  & 0.11 & 0.02 &  NC  &  NC  & 1.16  & 0.24  & 0.14  & 0.06  &  NC   &  NC   &  NC   &  NC   &  NC   & 1.10 \\
\hline 
S250           &   NC  & 0.61 & 0.02 &  NC  &  NC  &  NC   & 1.15  & 0.21  & 0.06  &  NC   &  NC   &  NC   &  NC   &  NC   &  NC  \\
S350           &   NC  & 6.08 & 0.08 &  NC  &  NC  &  NC   &  NC   & 1.15  & 0.14  &  NC   &  NC   &  NC   &  NC   &  NC   &  NC  \\
S500           &   NC  &  NC  & 0.47 &  NC  &  NC  &  NC   &  NC   &  NC   & 1.14  &  NC   &  NC   &  NC   &  NC   &  NC   &  NC  \\
\hline 
FUV              &  0.15 & 0.00 & 0.00 & 0.10 & 0.01 & 0.00  & 0.00  & 0.00  & 0.00  & 1.18  & 0.30  & 0.21  & 0.11  & 0.07  & 0.03 \\
NUV              &  0.33 & 0.00 & 0.00 & 0.47 & 0.07 & 0.00  & 0.00  & 0.00  & 0.00  & 3.35  & 1.18  & 0.60  & 0.26  & 0.12  & 0.03 \\
\hline 
W3.4            &  0.60 & 0.00 & 0.00 & 1.05 & 0.21 & 0.00  & 0.01  & 0.00  & 0.00  & 5.02  & 2.37  & 1.18  & 0.53  & 0.29  & 0.01 \\
W4.6            &  1.27 & 0.00 & 0.00 & 2.19 & 0.56 & 0.00  & 0.01  & 0.00  & 0.00  &  NC   & 4.29  & 2.43  & 1.18  & 0.83  & 0.04 \\
W11.6           &  2.06 & 0.02 & 0.00 & 3.14 & 1.14 & 0.22  & 0.07  & 0.02  & 0.00  &  NC   & 5.70  & 3.52  & 1.95  & 1.17  & 0.35 \\
W22.1           &   NC  & 0.39 & 0.01 &  NC  &  NC  & 1.65  & 0.60  & 0.28  & 0.04  &  NC   &  NC   &  NC   &  NC   &  NC   & 1.16 \\
\hline \hline 
\multicolumn{16}{l}{Notes.-- $W_{-} =$ is the integral of the negative values for each kernel [see eq. (\ref{eq:W})].} \\
\multicolumn{16}{l}{We are abbreviating I, M, P, S, GAL, and W }\\
\multicolumn{16}{l}{for IRAC, MIPS, PACS, SPIRE, GALEX, and WISE respectively.}\\
\multicolumn{16}{l}{NC stands for not computed: the kernel performance would be too poor.}\\
\end{tabular} 
\end{table*} 
 
Essentially, there are two sources of $W_{-}$: oscillations due to the filter $f_A$ and the need to remove energy from some region to relocate to another region (when the target $\mPSF_B$ is narrower than $\mPSF_A$ or has less energy in some annuli).

The kernels between cameras with similar FWHM also have oscillations. 
Using a softer filter $f_A$ would reduce the oscillatory behavior, giving smaller values of $W_{-}$, but would also produce worse matched PSFs (larger values of $D$). 
The particular form of filter $f_A$ used in the present work is a good compromise between having good PSF matching and moderate oscillatory behavior.

\begin{figure}[h] 
\centering 
\includegraphics[width=8.6cm,height = 8.0cm,clip=true,trim=0.0cm 0.0cm 0.0cm 1.2cm]
{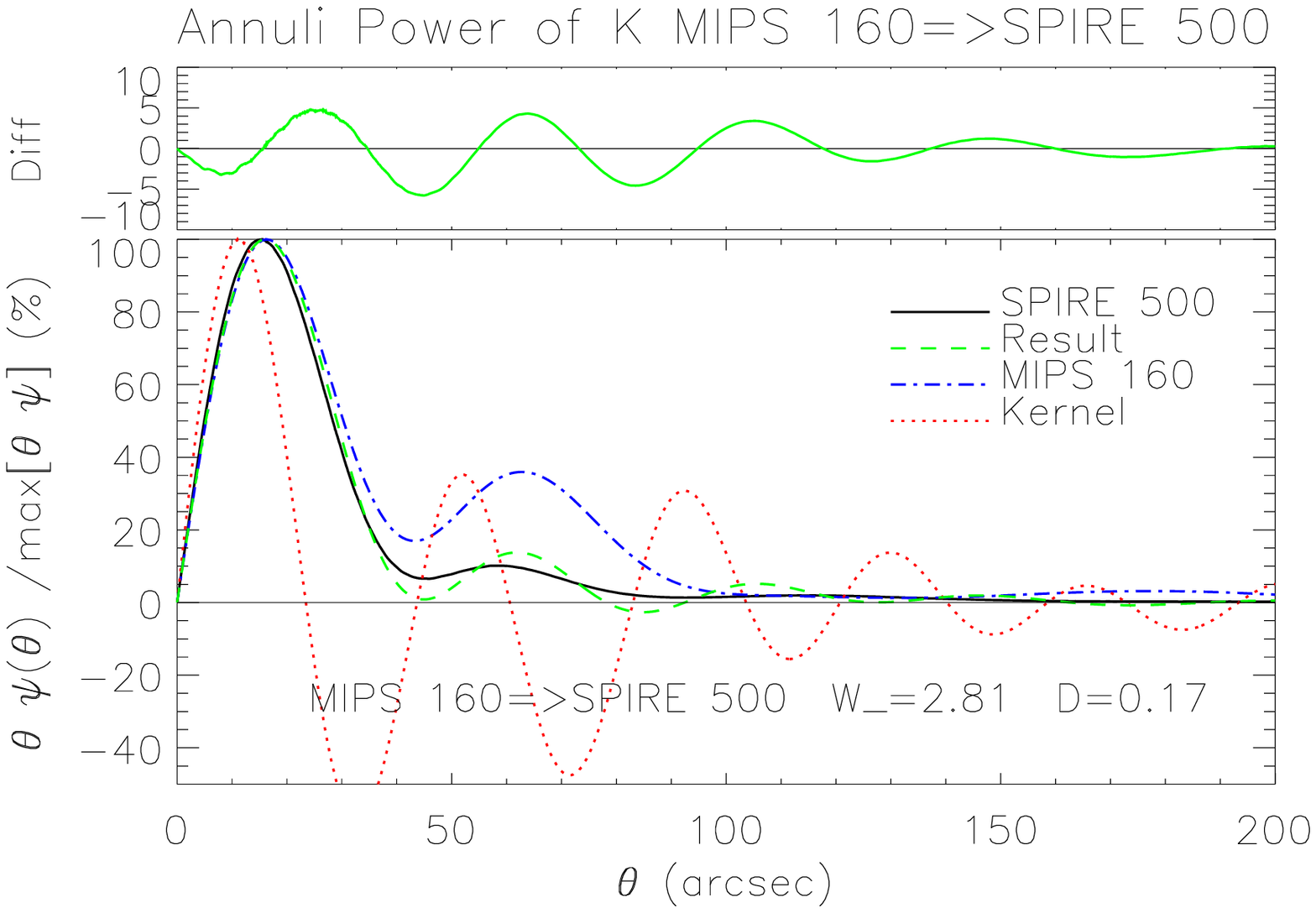} 
\caption{Performance of the kernel $K\{\mathrm{M}160 \Rightarrow\mathrm{S}500\}$.
With $W_{-}=2.81$, convolution of MIPS160\um images into SPIRE500\um resolution is risky and not recommended. 
The convolved PSF differs from the target PSF by up to 6\%.
See the electronic edition of the PASP for a color version of this figure.}
\label{fig:perf4}
\end{figure}

In Figure \ref{fig:perf4} we analyze a kernel of particular interest: $K\{\mathrm{M}160 \Rightarrow\mathrm{S}500\}$.
Despite both cameras having similar FWHM ($\mathrm{FWHM}_{\mathrm{M}160} / \mathrm{FWHM}_{\mathrm{S}500} = 1.07$), their extended wings are very different.
The kernel $K\{\mathrm{M}160 \Rightarrow\mathrm{S}500\}$ is particularly badly behaved, with large negative excursions, having  $W_{-}=2.81$ and $D=0.17$. 
In a convolution of MIPS 160\mic images of NGC 6946, some bright point sources generated regions with negative flux around them. 
We do not recommend using the kernel $K\{\mathrm{M}160 \Rightarrow\mathrm{S}500\}$: if MIPS160\um and SPIRE500\um images need to be combined, we recommend using the PSF of MIPS 160 \mic ( $K\{\mathrm{S}500 \Rightarrow\mathrm{M}160\}$ has $W_{-}\approx 0.47 $ and $D\approx 0.042 $) or some Gaussian PSF compatible with MIPS 160 \mic, such as a Gaussian with FWHM=64\asec (see \S 6).

\begin{figure}[h] 
\centering 
\includegraphics[width=8.6cm,height = 8.0cm,clip=true,trim=0.0cm 0.0cm 0.0cm 1.2cm]
{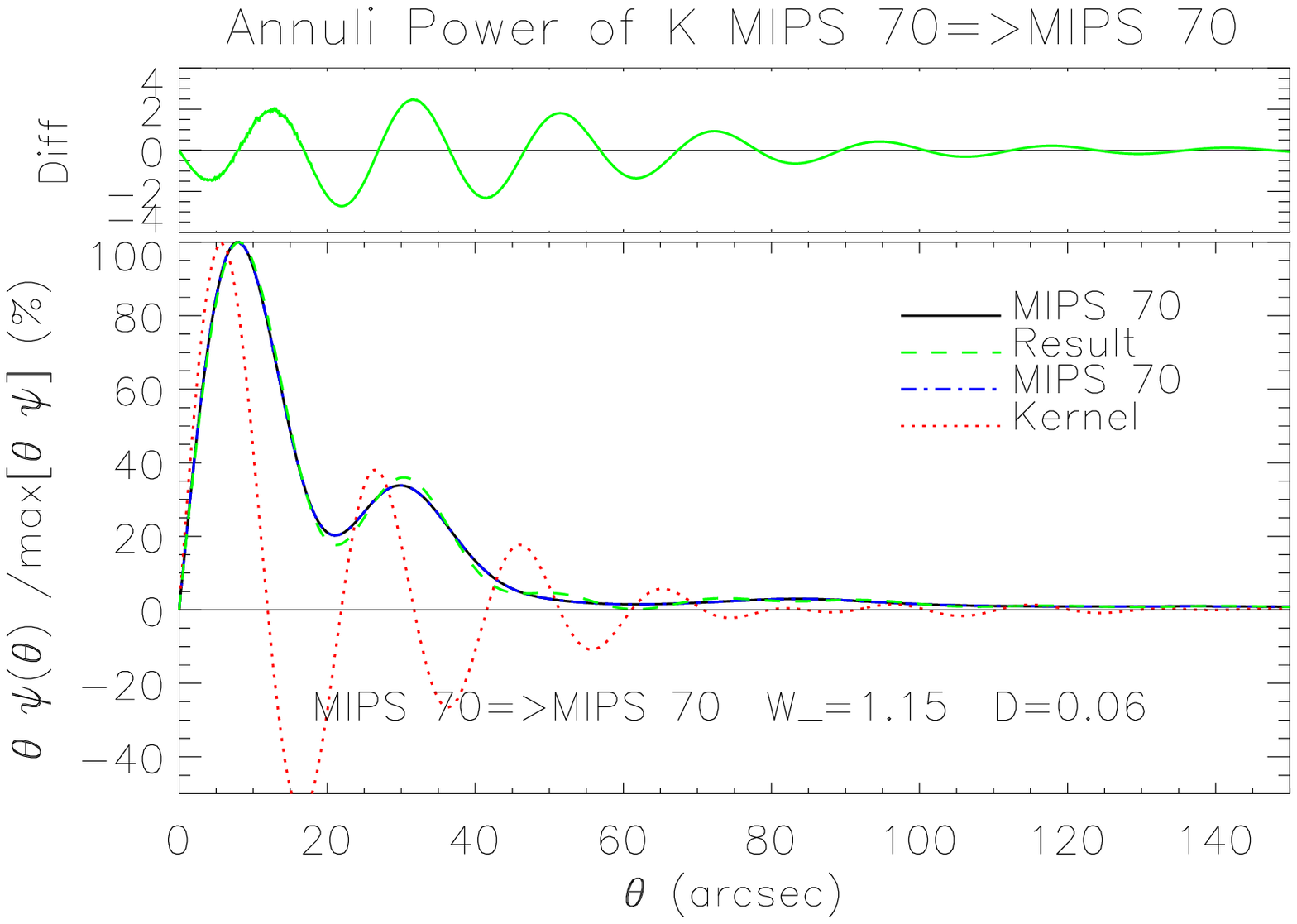} \\
\caption{Performance of the self- kernel $K\{\mathrm{M}70 \Rightarrow\mathrm{M}70\}$.
All of the self- kernels have $W_{-}=1.16$, and reproduce the original PSF to within 3\%.
See the electronic edition of the PASP for a color version of this figure.}
\label{fig:perf5}
\end{figure}

Finally, in Figure \ref{fig:perf5} we analyze the kernel $K\{\mathrm{M}70 \Rightarrow\mathrm{M}70\}$.
This kernel is essentially $FT(f_{\mathrm{M}70})$, and it illustrates the effect of all the kernel construction steps.
All the kernels of the form $K\{A \Rightarrow A\}$ are scaled versions of $K\{\mathrm{M}70 \Rightarrow\mathrm{M}70\}$, aside from small differences due to finite grids. All of the $K\{A \Rightarrow A\}$  kernels have $W_{-}\approx1.15$ and $D\approx0.06$.

Even though kernels $K\{A \Rightarrow A\}$ have $W_{-}=1.15$ and $W_{+}=2.15$, they do not amplify the noise that arises from astronomical sources.
The image of an astronomical point source will be $\mPSF_A$. 
When we convolve the camera with a kernel of the form $K\{A \Rightarrow A\}$, the image of a point source will still be very close to $\mPSF_A$, since $D(K\{A \Rightarrow A\})\sim0.06$. 
This implies that the noise coming from a field of unresolved astronomical background sources will not change significantly when we convolve the image with $K\{A \Rightarrow A\}$.
To verify this reasoning, we generate an image $S$ having independent Gaussian noise in each pixel in a very fine (0.2\asec) grid.
We convolve $S$ with $\mPSF_{\mathrm{M}70}$ to have a simulated observed image of the noise: $O=S\star \mPSF_{\mathrm{M}70}$. We further convolve $O$ with $K\{\mathrm{M}70 \Rightarrow\mathrm{M}70\}$: $C=O\star K\{\mathrm{M}70 \Rightarrow\mathrm{M}70\}$.
We found that $\mid 1-\langle C^2\rangle / \langle O^2\rangle \mid  \approx 10^{-3}$.
While uncorrelated noise is not amplified by a self-kernel $K\{A \Rightarrow A\}$, imprecise characterization of the PSFs and camera artifacts can be amplified by kernels with large $W_{-}$ values.

There is no single number that captures all of the characteristics of a convolution kernel, but we find that $W_{-}$ serves as a good figure of merit. Based on experimentation with various kernels, we regard kernels with $W_{-} \le 0.3$ to be very safe.
Kernels with $W_{-} \approx 0.5$ also appear to be quite safe. A kernel with $W_{-}\approx 1$ is somewhat aggressive in moving power around, but remember that self-kernels $K\{A \Rightarrow A\}$ also have $W_{-}\approx 1.15$.
We consider kernels with $W_{-}\approx 1$ to be reasonable to use, with inspection of the before and after images recommended in regions with large gradients or bright sources. 
We recommend against using convolution kernels with $W_{-}>1.2$, as these are, in effect, attempts to deconvolve the image to higher resolution, with attendant risks.


\section{Gaussian PSFs}

Gaussian PSFs are commonly used in radio astronomy. 
It is desirable for radio telescopes to have PSFs without extended structure to avoid sidelobe contamination. 
The illumination pattern of a single dish is often designed to return an approximately Gaussian PSF.\footnote{Interferometric arrays have complicated sidelobe responses and would not resemble Gaussian PSFs.}
A Gaussian PSF lacks extended wings; the fraction of the power outside radius $\theta$  is $ \exp \left({-\theta^2 / 2\sigma^2}\right)$. 
Because real instrumental PSFs do not fall off so rapidly, a convolution kernel $K$ going from a real $\mPSF_A$ to a Gaussian $\mPSF_B$ with similar FWHM must  ``move'' power from the wings of $\mPSF_A$ to the core of $\mPSF_B$.

For a given instrumental PSF $\mPSF_A$, the optimal Gaussian PSF $\mPSF_B$ will be such that the $\mathrm{FWHM}_{B}$ will be close to $\mathrm{FWHM}_{A}$, with only mild filtering by the function $f_A$ and with $W_{-}$ not too large. 

In order to determine an optimal Gaussian FWHM for a camera $A$, we compute convolution kernels from  $A$ into Gaussian PSFs with FWHM in a range of possible values.
For each candidate FWHM, we evaluate $W_{-}$.
We provide three possible FWHM. 
The first FWHM is obtained by requiring that $W_{-} \sim 0.3$, giving a conservative (very safe) kernel that does not seek to move too much energy from the wings into the main Gaussian core, at the cost of having a larger FWHM (i.e., lower resolution). 
The second FWHM has $W_{-} \sim 0.5$, and we consider it to be a good (moderate) Gaussian FWHM to use.
The third FWHM has $W_{-} \sim 1.0$. 
Because this kernel is somewhat ``aggressive" in moving energy from the PSF wings into the Gaussian core, it should be used with care, inspecting that the convolved images do not have any induced artifact. Table \ref{tab_Ker_Gauss} gives the FWHM for three such Gaussian target PSFs for the MIPS, PACS, and SPIRE cameras.

\begin{table*}[h] 
\caption{Gaussian FWHM Suitable for MIPS, PACS, and SPIRE Cameras} 
\label{tab_Ker_Gauss}
\centering 
\begin{tabular}{|l|c|cc|cc|cc|} 
\hline \hline 
       &  Actual & \multicolumn{2}{|c|}{Aggressive Gaussian}     & \multicolumn{2}{|c|}{Moderate Gaussian}        &  \multicolumn{2}{|c|}{Very safe Gaussian}       \\
Camera &  FWHM   & \multicolumn{2}{|c|}{with $W_{-}\approx 1.0$} & \multicolumn{2}{|c|}{with $W_{-}\approx  0.5$} &  \multicolumn{2}{|c|}{with $W_{-}\approx  0.3$} \\
       & (\asec) & FWHM (\asec) & $W_{-}$                        & FWHM (\asec) & $W_{-}$                         & FWHM (\asec) & $W_{-}$                          \\
\hline
\hline
MIPS 24\mic   &  6.5 &  8.0 & 1.00 & 11.0 & 0.49 & 13.0 & 0.30 \\
MIPS 70\mic   & 18.7 & 22.0 & 1.01 & 30.0 & 0.51 & 37.0 & 0.30 \\
MIPS 160\mic  & 38.8 & 46.0 & 1.01 & 64.0 & 0.50 & 76.0 & 0.30 \\
\hline
PACS 70\mic   &  5.8 &  6.5 & 0.84 &  8.0 & 0.48 & 10.5 & 0.31 \\
PACS 100\mic  &  7.1 &  7.5 & 1.10 &  9.0 & 0.52 & 12.5 & 0.31 \\
PACS 160\mic  & 11.2 & 12.0 & 1.05 & 14 & 0.50 & 18.0 & 0.33 \\
\hline
SPIRE 250\mic & 18.2 & 19.0 & 1.05 & 21.0 & 0.44 & 22.0 & 0.30 \\
SPIRE 350\mic & 25.0 & 26.0 & 0.98 & 28.0 & 0.50 & 30.0 & 0.27 \\
SPIRE 500\mic & 36.4 & 38.0 & 0.96 & 41.0 & 0.48 & 43.0 & 0.30 \\
\hline \hline 
\end{tabular}
\end{table*}

\begin{figure*}[h] 
\centering 
\includegraphics[width=16cm,height = 20cm,clip=true,trim=0.0cm 0.0cm 0.0cm 0.0cm]
{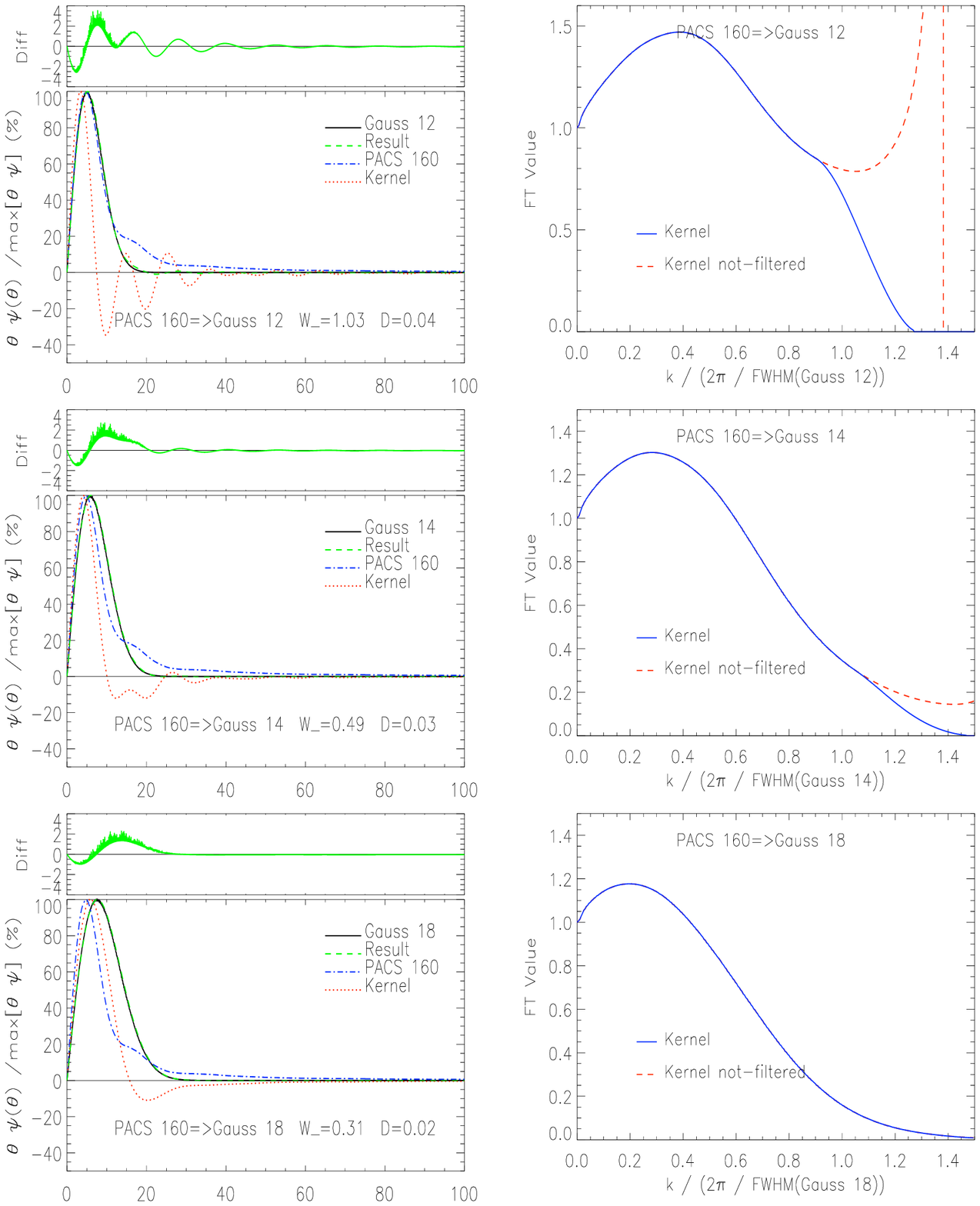} 
\caption{Performance of the kernels $K\{\mathrm{P}160 \Rightarrow\mathrm{Gaussian}\}$ (left panel), and their $FT$ (right panel). 
We show the kernel Fourier transform $FT(K)$ (equation (\ref{eq:fil_FT})), and also the unfiltered kernel Fourier transform ($= FT_\phi(\mPSF_{B}) / FT_\phi(\mPSF_A)$).
The filter $f_{\mathrm{P}160}$ 
has little impact on $K\{\mathrm{P}160 \Rightarrow\mathrm{Gaussian}18\masec\}$ ($W_{-}=0.33$),
has moderate effect on $K\{\mathrm{P}160 \Rightarrow\mathrm{Gaussian}14\masec\}$ ($W_{-}=0.50$), 
and large effect on $K\{\mathrm{P}160  \Rightarrow\mathrm{Gaussian}12\masec\}$($W_{-}=1.05$).
See the electronic edition of the PASP for a color version of this figure.}
\label{im_GaussPacs160}
\end{figure*}

As an example, Figure \ref{im_GaussPacs160} shows the performance of the kernels for the PACS 160\mic camera going into Gaussian PSFs with FWHM = 12\asec (aggressive,  $W_{-}=1.05\,D=0.05$), 14\asec (moderate, $W_{-}=0.50,\,D=0.02$), and 18\asec (very safe, $W_{-}=0.33,\,D=0.01$). 
The left panels are the performance analyses, similar to those of Figure \ref{fig:perf2} - \ref{fig:perf5}. 
The right panels show the kernel Fourier transform $FT(K)$ (equation (\ref{eq:fil_FT}). 
In the right panel we include the unfiltered kernel Fourier transform ($= FT_\phi(\mPSF_{\mathrm{Gauss}})/FT_\phi(\mPSF_{\mathrm{P}160)}$) for comparison (dotted line). 
We observe that the filtering is only important in the Gaussian kernels with (small)  FWHM = 12\asec and 14\asec .

\clearpage


\section{Usage of the Kernels}

The kernels $K\{A \Rightarrow B\}$ computed here are given on a 0.20\asec grid.
Before performing an image convolution, the kernel $K\{A \Rightarrow B\}$ should be resampled onto a grid with the same pixel size as the original image $I_A$.
The resampled kernels should be centered (to avoid shifts in the image) and normalized so that $\int\int {K\{A \Rightarrow B\}(x,y)} dxdy=1$ to ensure flux conservation.
The flux in the image to be convolved should be in surface brightness units.
After convolving the image $I_A$ with the kernel $K\{A \Rightarrow B\}$, the resulting image will be expressed in the original image grid and original surface brightness units, but with PSF $\mPSF_B$.

Table \ref{tab_Ker_W} also summarizes the kernels available.
For each camera $A$ we construct all the kernels $K\{A \Rightarrow B\}$ with $ \mathrm{FWHM}_{B} \ge \mathrm{FWHM}_{A}/1.35$ (i.e., the kernels that degrade the resolution or sharpen it up to 35$\%$) plus the self-kernels $K\{A \Rightarrow A\}$. 
Kernels with $ \mathrm{FWHM}_{A} \gtrsim \mathrm{FWHM}_{B}$ (that have larger $W_{-}$ values) tend to perform poorly and should be used with care. We do not recommend using any kernel with  $W_{-}\gtsim 1.2 $ 

\begin{figure*}
\centering 
\includegraphics[width=16cm,height = 16cm,clip=true,trim=0.0cm 0.0cm 0.0cm 0.0cm]
{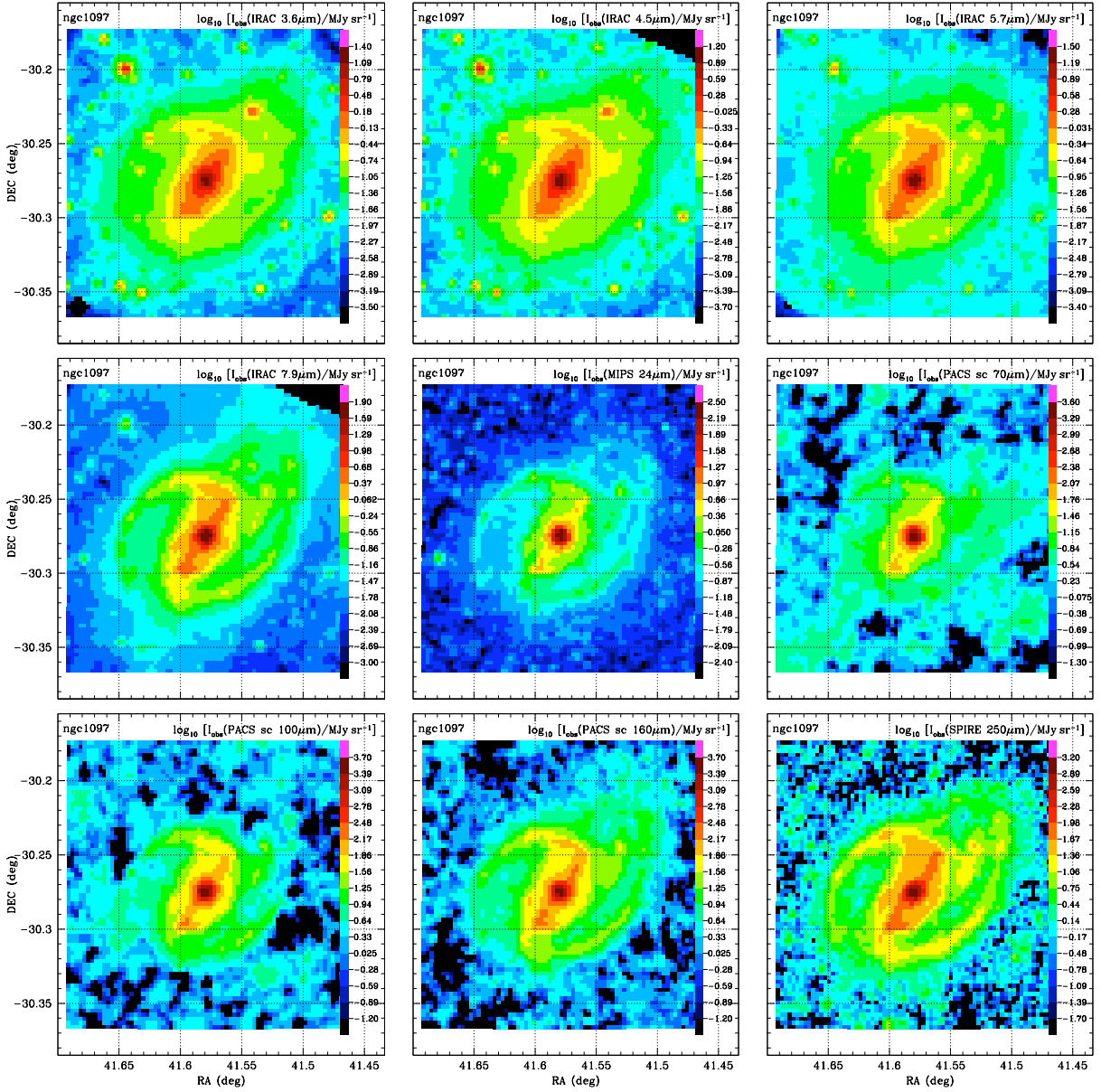} 
\caption{Spitzer and Herschel images of NGC 1097 convolved to a SPIRE 250\mic PSF. The SPIRE 250\mic camera was convolved with the kernel  $K\{\mathrm{S}250 \Rightarrow\mathrm{S}250 \}$. The color bar has the same dynamic range $(10^{4.9})$ for all images.
See the electronic edition of the PASP for a color version of this figure.}
\label{im_n1097_S250}
\end{figure*}

As an example of the performance of the kernels applied to real (noisy) images, in Figure \ref{im_n1097_S250} we show the result of convolving images of the barred spiral galaxy NGC 1097 (after subtraction of a ``tilted-plane'' background from each image)into the SPIRE 250 \mic PSF. 
The PACS images have been reduced using the Scanamorphos pipeline \citep{Roussel_2011}.
Visual inspection of the images in Figure \ref{im_n1097_S250} shows them to be very similar in morphology; the convolution does not appear to have introduced any noticeable artifact. 

\begin{figure*}
\centering 
\includegraphics[width=16cm,height = 10.5cm,clip=true,trim=0.0cm 0.0cm 0.0cm 0.0cm]
{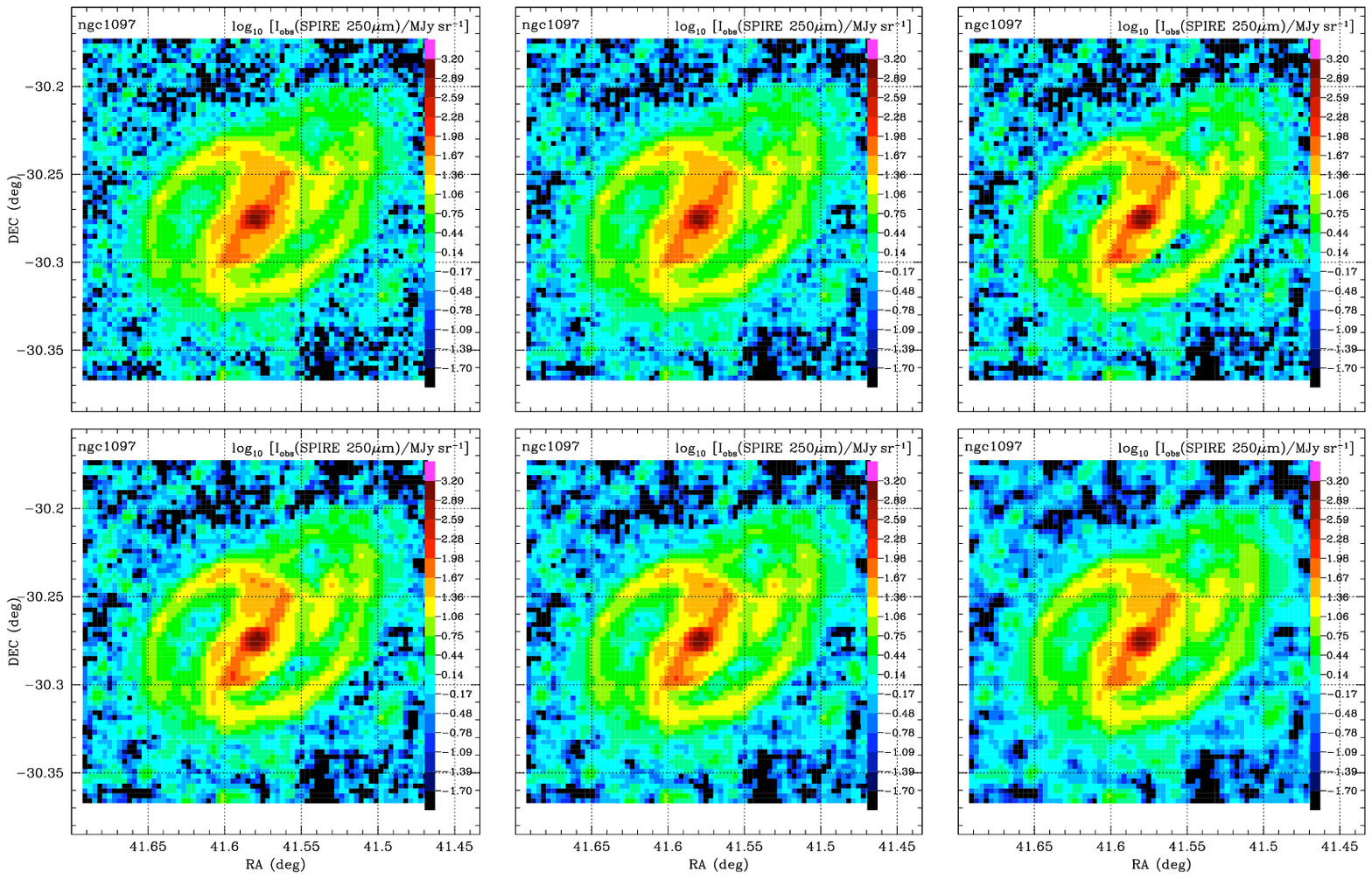} 
\caption{SPIRE 250\mic image of NGC 1097. Top row left: Original SPIRE image. Center: Image convolved with $K\{\mathrm{S}250 \Rightarrow\mathrm{S}250\}$. Right: image convolved into an extremely aggressive Gaussian PSF with FWHM = 18\asec ($W_{-}=1.47$). Bottom row: Image convolved into suitable Gaussian PSFs. Left: (aggressive) FWHM = 19\asec ($W_{-}=1.05$). Center: (moderate) FWHM = 21\asec ($W_{-}=0.44$). Right: (very safe) FWHM = 22\asec ($W_{-}=0.30$). 
All the images have the same color bar. 
See the electronic edition of the PASP for a color version of this figure.}
\label{im_n1097_various}
\end{figure*}

Figure \ref{im_n1097_various} shows the results of convolving the SPIRE 250 \mic image into different PSFs. 
The top row (left) shows the original\footnote
{By ``original'' image we refer to the SPIRE 250\mic image delivered by the HIPE pipeline, with subsequent subtraction of a ``tilted-plane'' background. This is the original image that is convolved to produce the other images in Fig. \ref{im_n1097_various}.} image,  (center) the image convolved with $K\{\mathrm{S}250 \Rightarrow\mathrm{S}250\}$, and (right) the image convolved with a very aggressive kernel into a Gaussian PSF with FWHM=18\asec ($W_{-}=1.47$). 
The bottom row shows the image convolved to the recommended Gaussian PSFs: (left) FWHM = 19\asec ($W_{-}=1.05$), (center) FWHM = 21\asec ($W_{-}=0.44$), and (right) FWHM = 22\asec ($W_{-}=0.30$).  
As discussed in \S 5, we recommend against using kernels with $W_{-}>1.2$.
Visual inspection of the upper-right image in fact shows artifacts where the kernel (with $W_{-}=1.47$) has moved too much power out of some interarm pixels, which have been brought down to unreasonably low intensities.
In the convolutions to the suitable Gaussian PSFs (bottom row in Fig. \ref{im_n1097_various}) energy is moved from the interarm regions into the bright nucleus and spiral arms, but the intensity levels in the interarm regions seem reasonable.
The power that is removed from the interarm regions was presumably originally power from the arms that was transferred by the wings of the SPIRE 250\um PSF.

Using the kernels described in the current work, \citet{Aniano+Draine+KINGFISH_2011a} studied resolved dust modeling for NGC 628 and NGC 6946, two galaxies in the KINGFISH galaxy sample \citep{Kennicutt+KINGFISH_2011}, using images obtained with Spitzer and Herschel Space Observatory.

\section{Summary}

We present the construction and analysis of convolution kernels, for transforming images into a common PSF. 
They allow generation of a multiwavelength image cube with a common PSF, preserving the colors of the regions imaged.

We generate and make available a library of convolution kernels for the cameras of the Spitzer, Herschel Space Observatory, GALEX, WISE, ground-based telescopes, and Gaussian PSFs. 
All the kernels are constructed with the best PSF characterizations available, approximated by rotationally symmetric functions.
Deviations of the actual PSF from circular symmetry are characterized by an asymmetry parameter $g$, given in Table \ref{tab_PSF_info}.
Table \ref{tab_Ker_W} summarizes the kernels available and their negative integral $W_{-}$, a measure of their performance.
We recommend using only kernels with $W_{-}\lesssim 1.2$.
In Table \ref{tab_Ker_Gauss} we give a set of optimal Gaussian FWHM that are compatible with MIPS, PACS, and SPIRE cameras.

All the kernels, IDL routines to use the kernels and IDL routines to make new kernels, along with detailed analysis of the generated kernels, are publicly available.\footnote{See http://www.astro.princeton.edu/$\sim$draine/Kernels.html. Kernels for additional cameras and updates will be included when new PSF characterizations become available.}

We thank Roc Cutri and Edward Wright for their help providing the Wide-field Infrared Survey Explorer point-spread functions (PSFs); Markus Nielbock for his help providing the Photocamera Array Camera and Spectrometer for Herschel PSFs; and Richard Bamler, James Gunn, Robert Lupton and the anonymous referee for helpful suggestions.

See the electronic edition of the PASP for a color version of this figure.

This research was supported in part by NSF grant AST-1008570 and JPL grant 1373687.

\bibliography{btdrefs}

\end{document}